\documentclass[12pt]{article}
\usepackage{latexsym,amsmath,amssymb,bm,cite}
\usepackage{graphicx}
\usepackage{color}
\newcommand{\be}{\begin{equation}}
\newcommand{\ee}{\end{equation}}
\newcommand{\bea}{\begin{eqnarray}}
\newcommand{\eea}{\end{eqnarray}}
\newcommand{\bem}{\begin{multline}}
\newcommand{\eem}{\end{multline}}
\newcommand{\beg}{\begin{gather}}
\newcommand{\eeg}{\end{gather}}

\newcommand{\as}{\alpha_s}

\def\eq#1{{Eq.~(\ref{#1})}}
\def\fig#1{{Fig.~\ref{#1}}}
\newcommand{\ben}{\begin{eqnarray*}}
\newcommand{\een}{\end{eqnarray*}}

\newcommand{\am}{\alpha_\mu}

\begin{document}
\noindent{\Large\bf Quark loop contribution to BFKL evolution:}\\[.2cm]
\noindent{\Large\bf Running coupling and leading-$N_f$
     NLO intercept } 
\\
\vspace{0.5cm}
\\
{\large
    Yuri V.\ Kovchegov$^1$ and Heribert Weigert$^2$
}\\\vspace{0.2cm}\\
{\small $^1$Department of Physics, The Ohio State University, 
Columbus, OH 43210, USA}
\\
{\small $^2$Fakult\"{a}t f\"{u}r Physik, Universit\"{a}t Bielefeld,
  D-33615, Bielefeld, Germany}\\




\vspace{.2cm}
\noindent\begin{center}
\begin{minipage}{.94\textwidth}
  {\sf We study the sea quark contribution to the BFKL kernel in the
    framework of Mueller's dipole model using the results of our
    earlier calculation.  We first obtain the BFKL equation with the
    running coupling constant. We observe that the ``triumvirate''
    structure of the running coupling found previously for non-linear
    evolution equations is preserved for the BFKL equation.  In fact,
    we rederive the equation conjectured by Levin and by Braun, albeit
    for the unintegrated gluon distribution with a slightly
    unconventional normalization. We obtain the leading-$N_f$
    contribution to the NLO BFKL kernel in transverse momentum space
    and use it to calculate the leading-$N_f$ contribution to the NLO
    BFKL pomeron intercept for the unintegrated gluon distribution.
    Our result agrees with the well-known results of Camici and
    Ciafaloni and of Fadin and Lipatov.  We show how to translate this
    intercept to the case of the quark dipole scattering amplitude and
    find that it maps onto the expression found by Balitsky.  }
\end{minipage}
\end{center}
\vspace{1cm}





\section{Introduction}

Understanding the NLO corrections to the Balitsky-Fadin-Kuraev-Lipatov
(BFKL) \cite{Kuraev:1977fs,Bal-Lip}, the Balitsky-Kovchegov (BK)
\cite{Balitsky:1996ub,
  Balitsky:1997mk,Balitsky:1998ya,Kovchegov:1999yj, Kovchegov:1999ua}
and the Jalilian-Marian--Iancu--McLerran--Weigert--Leonidov--Kovner
(JIMWLK) \cite{Jalilian-Marian:1997jx, Jalilian-Marian:1997gr,
  Jalilian-Marian:1997dw, Jalilian-Marian:1998cb, Kovner:2000pt,
  Weigert:2000gi, Iancu:2000hn,Ferreiro:2001qy} evolution equations is
important to have their numerical predictions under control. Improving
the precision of the solutions of BK and JIMWLK evolution equations
has far-reaching consequences for our understanding of the QCD physics
at HERA, RHIC, LHC colliders and at the proposed eRHIC/EIC collider.
It would sharpen the predictions of the saturation/Color Glass
Condensate (CGC) physics
\cite{Gribov:1981ac,Mueller:1986wy,McLerran:1994vd,McLerran:1994ka,
  McLerran:1994ni,Kovchegov:1996ty,Kovchegov:1997pc,Jalilian-Marian:1997xn,
  Jalilian-Marian:1997jx, Jalilian-Marian:1997gr,
  Jalilian-Marian:1997dw, Jalilian-Marian:1998cb, Kovner:2000pt,
  Weigert:2000gi, Iancu:2000hn,Ferreiro:2001qy,Kovchegov:1999yj,
  Kovchegov:1999ua, Balitsky:1996ub, Balitsky:1997mk,
  Balitsky:1998ya,Iancu:2003xm,Weigert:2005us,Jalilian-Marian:2005jf}
for such observables as total and diffractive cross sections, particle
yields and spectra, and particle correlations. This would move the CGC
physics from the qualitative phase to the domain of quantitative
predictions.

Recently there has been much progress in calculating the running
coupling corrections for the JIMWLK \cite{Jalilian-Marian:1997jx,
  Jalilian-Marian:1997gr, Jalilian-Marian:1997dw,
  Jalilian-Marian:1998cb, Kovner:2000pt, Weigert:2000gi,
  Iancu:2000hn,Ferreiro:2001qy} and the BK \cite{Balitsky:1996ub,
  Balitsky:1997mk, Balitsky:1998ya,Kovchegov:1999yj, Kovchegov:1999ua}
evolution equations \cite{HW-YK:2006,Gardi:2006,Bal:2006}. The idea
behind the calculation of running coupling corrections in
\cite{HW-YK:2006,Bal:2006} was to calculate the quark (leading-$N_f$)
contribution to the running of the coupling and recover the
leading-order QCD beta-function by replacing $N_f \rightarrow - 6 \,
\pi \, \beta_2$ in the result. ($\beta_2$ is given by \eq{beta2}
below.)

The leading-$N_f$ contribution to the running of the coupling was
calculated in \cite{HW-YK:2006,Bal:2006} by inserting quark loops to
all orders in the gluon line emitted per one step of the JIMWLK or BK
evolution. If one limits the expansion to just one quark loop, one
would then obtain the leading-$N_f$ contribution to the NLO JIMWLK and
BK kernels. Since the linearized versions of the JIMWLK and BK
equations give the BFKL equation \cite{Kuraev:1977fs,Bal-Lip}, the
results of \cite{HW-YK:2006,Bal:2006} contain all the ingredients
needed to calculate the leading-$N_f$ NLO correction to the BFKL
pomeron intercept. To obtain the leading-$N_f$ NLO BFKL equation from
the NLO JIMWLK and BK equations with running coupling corrections one
should simply keep only the linear terms in the latter equations.


It is a well-known problem that the NLO correction to the BFKL pomeron
intercept calculated in
\cite{Fadin:1998py,Fadin:1995xg,Fadin:1996zv,Ciafaloni:1998gs,Camici:1997ta}
is large and negative (for other problems of NLO BFKL see
\cite{Kovchegov:1998ae,Ross:1998xw}). The largeness of the correction
was argued to be due to collinear singularities
\cite{Ciafaloni:1999yw,Salam:1998tj}. Since the gluon saturation
effects in both the BK and JIMWLK evolutions cut off the contributions
from the dangerous infrared region, there is a hope that NLO
corrections to the BK and JIMWLK equations would be numerically small.
The work presented below is the first step aimed at verification of
the above hypothesis in the sense that it explicitly demonstrates the
relationship between the JIMWLK/BK and the BFKL equations beyond
leading order.


In ~\cite{Bal:2006} the leading-$N_f$ contribution to the NLO
correction to the BFKL pomeron intercept was calculated for the
scattering amplitude of a dipole on a target. The calculation was
performed in transverse coordinate space leading to an intercept
different from the one previously obtained by Camici and Ciafaloni in
\cite{Camici:1997ta} and by Fadin and Lipatov in \cite{Fadin:1998py}
using conventional momentum space perturbation theory.  Below we
address the origin of this discrepancy by calculating the
leading-$N_f$ contribution to the NLO BFKL intercept for a different
observable: we will deal with the unintegrated gluon distribution,
defined below by Eqs. (\ref{phi}) and (\ref{FT}). We show that the
leading-$N_f$ contribution to the NLO BFKL intercept for the
unintegrated gluon distribution obtained below in fact agrees with the
results of \cite{Camici:1997ta,Fadin:1998py}!


In \cite{HW-YK:2006} the running coupling corrections to the JIMWLK/BK
evolution equations were calculated to all orders. This allows us to
derive the BFKL equation including the running coupling corrections.
We will do this below by keeping only the linear term in the running
coupling JIMWLK/BK equations obtained in \cite{HW-YK:2006}. The
result, shown in \eq{rc_BFKL}, gives a first-ever derivation of the
BFKL equation with the running coupling corrections resummed to all
orders. While \eq{rc_BFKL} is written for the unintegrated gluon
distribution, a slight change in the normalization of that quantity
shown in \eq{tphi} leads to \eq{rc_BFKL_Genya}. \eq{rc_BFKL_Genya} is
exactly the equation which was conjectured by Braun and by Levin in
\cite{Braun:1994mw,Levin:1994di} by {\sl ad hoc} postulating the
bootstrap condition to remain valid even when the running coupling
corrections are included. In this work we present a first-principle
derivation of the equation conjectured by Braun and by Levin,
confirming its validity for a quantity slightly different from the
conventional unintegrated gluon distribution. A partial validity check
of \eq{rc_BFKL_Genya} was previously performed in \cite{BV1,BV2}:
there the leading-$N_f$ contribution to the NLO BFKL intercept was
calculated after expanding \eq{rc_BFKL_Genya} to the next-to-leading
order and was found to be exactly the same as obtained in
\cite{Camici:1997ta,Fadin:1998py}.

The paper is structured as follows. In Sect. \ref{DipBFKL} we
establish the calculational framework with a leading order comparison
of JIMWLK/BK on the one hand and BFKL on the other hand. We start with
the Mueller's dipole model
\cite{Mueller:1994rr,Mueller:1994jq,Mueller:1995gb,Chen:1995pa}
analogue of the BFKL equation. We present the subtleties involved in
performing the Fourier-transform in into transverse momentum space and
derive the resulting equation for the unintegrated gluon distribution
function. We show that this evolution equation is indeed equivalent to
the standard BFKL equation by calculating the eigenfunctions and
eigenvalues of the kernel.

In Sect. \ref{momNLO} we use the results of \cite{HW-YK:2006} to
construct the leading-$N_f$ NLO correction to the BFKL kernel for the
evolution of the unintegrated gluon distribution in the large-$N_c$
approximation. The evolution equation combining the LO BFKL kernel and
leading-$N_f$ NLO correction to the kernel is given in \eq{NLO_BFKL}.

In Sect. \ref{rcBFKL} we construct the BFKL equation with the running
coupling corrections included to all orders. The resulting equation
for the unintegrated gluon distribution is shown in (\ref{rc_BFKL}).
This is the first main result of our paper. We show that by a simple
substitution \eq{rc_BFKL} can be recast into a form given by
\eq{rc_BFKL_Genya}, which was conjectured by Levin \cite{Levin:1994di}
and by Braun \cite{Braun:1994mw}, who postulated the validity of
bootstrap condition beyond the leading order. We compare our results
to the results of \cite{Levin:1994di,Braun:1994mw} in Sect.
\ref{nonfor} for the case of non-forward BFKL exchange. Again we show
that we can reproduce the equation conjectured in
\cite{Levin:1994di,Braun:1994mw} and clarify the physical quantity for
which that equation was written.

We proceed in Sect. \ref{NLOint} by calculating the leading-$N_f$
large-$N_c$ correction to the BFKL pomeron intercept, which is defined
by acting with the leading-$N_f$ NLO BFKL kernel on the eigenfunctions
of the LO BFKL kernel. Our result is given in \eq{int_gamma} and is in
complete agreement with the results of Camici and Ciafaloni
\cite{Camici:1997ta} and Fadin and Lipatov \cite{Fadin:1998py}. We
have thus verified the results of \cite{Camici:1997ta,Fadin:1998py}
using an independent method. This is the second main result of our
paper.

In Sect. \ref{Ian} we demonstrate that the difference between our
result and that of Balitsky \cite{Bal:2006} is solely due to the fact
that we are calculating evolution of different observables: while in
this paper we have been dealing with the unintegrated gluon
distribution, the intercept of \cite{Bal:2006} was calculated for the
dipole cross section. This difference demonstrates how different the
NLO BFKL intercept can be for different physical observables!


\section{The dipole model BFKL equation in transverse momentum space}
\label{DipBFKL}

\subsection{Fourier transform}

The conventional dipole model analogue of the BFKL equation
\cite{Bal-Lip,Kuraev:1977fs} reads
\cite{Mueller:1994rr,Mueller:1994jq,Mueller:1995gb,Chen:1995pa}
\begin{align}\label{BFKL}
  & \frac{\partial N ({\bm x}_{0}, {\bm x}_1, Y)}{\partial Y} \, = \,
  \frac{\as \, N_c}{2 \, \pi^2} \, \int d^2 x_2 \,
  \frac{x_{01}^2}{x_{20}^2 \, x_{21}^2} \, \left[ N ({\bm x}_{0}, {\bm
      x}_2, Y) + N ({\bm x}_{2}, {\bm x}_1, Y) - N ({\bm x}_{0}, {\bm
      x}_1, Y) \right],
\end{align}
where ${\bm x}_{ij} = {\bm x}_i - {\bm x}_j$ and $x_{ij} = |{\bm
  x}_{ij}|$. First, for simplicity, we assume that the forward
scattering amplitude of a dipole on a nucleus does not depend on the
impact parameter of the dipole and is also independent of the dipole's
orientation in the transverse plane:
\begin{align}\label{bindep}
  N ({\bm x}_{0}, {\bm x}_1, Y) \, = \, N (x_{01}, Y). 
\end{align}
The information contained in this simplified object is sufficient to compare
to the forward, angular--averaged BFKL equation in momentum space.

To transform \eq{BFKL} to transverse momentum space let
us define
\begin{align}\label{FT}
  N (x_{01}, Y) \, = \, \int \frac{d^2 k}{(2 \pi)^2} \, \left(1 - e^{i
      {\bm k} \cdot {\bm x}_{01}} \right) \, {\tilde N} (k,Y).
\end{align}
Indeed this is not the only way to transform \eq{BFKL} into momentum space:
see \cite{Kovchegov:1999ua} for an alternative. However, it appears that
\eq{FT} has the most straightforward application for the NLO kernel. 

We recall that $N$ has the interpretation of a dipole scattering cross
section, and as such, must vanish in the local limit: $\lim\limits_{{\bm
    x}_{0}\to{\bm x}_1} N ({\bm x}_{0}, {\bm x}_1, Y) = 0$; zero size dipoles
do not interact.

This can not be achieved by the using just a naive Fourier transformation, say
via the second term in~(\ref{FT}) alone. In fact, if one expands the generic
$N$ at low densities, one obtains
\begin{align}
\label{eq:N-low-density}
N ({\bm x}_{0}, {\bm x}_1, Y) \approx \frac12
\left(
 \parbox{2cm}{\includegraphics[width=2cm]{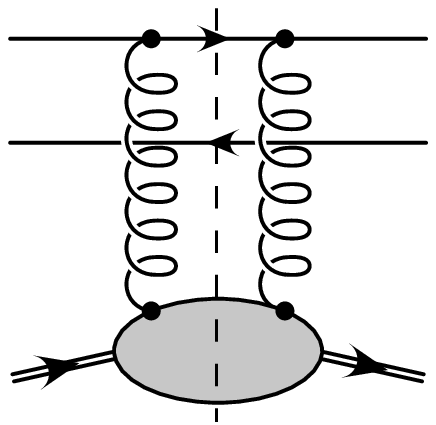}}
+\parbox{2cm}{\includegraphics[width=2cm]{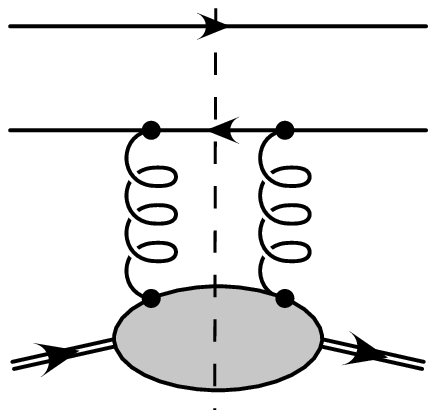}}
-\parbox{2cm}{\includegraphics[width=2cm]{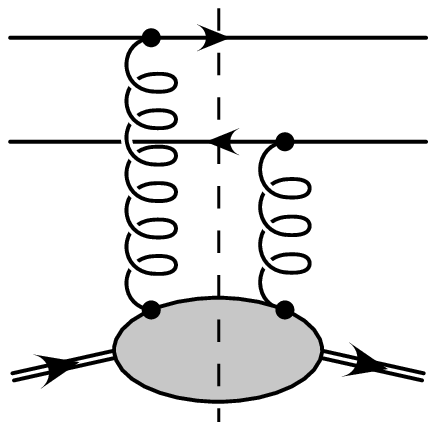}}
-\parbox{2cm}{\includegraphics[width=2cm]{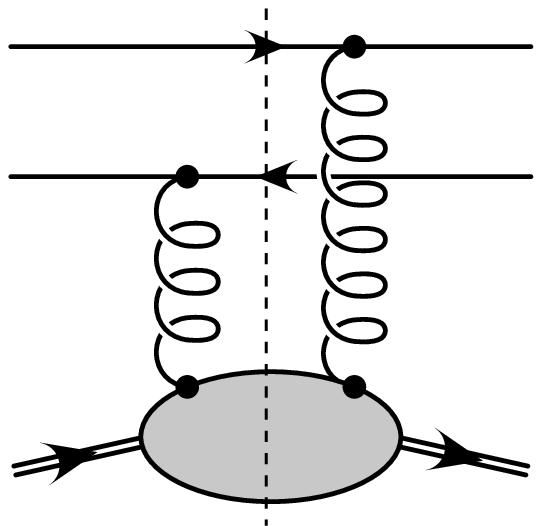}}
\right)
+\text{higher orders}
\end{align}
where the horizontal lines at ${\bm x}_{0}$ and ${\bm x}_1$ serve to mark the
transverse positions of the quark and antiquark. (Note that the complete
contribution in~\eqref{eq:N-low-density} is positive as it is an absolute
value of the difference of two amplitudes squared. Similarly the two first
diagrams are positive.) The blobs in the diagrams are related to the two point
correlator of the target fields (integrated over longitudinal positions). The
relative sign between the two types of contributions ensures that the local
limit vanishes. This property is preserved if we Fourier--transform the 4
contributions \emph{individually}. To this end we identify (thereby
restricting ourselves to the forward case)
\begin{align}
  \label{eq:tilde-N-k}
  \Tilde N (k, Y) = 
  \parbox{2cm}{\includegraphics[width=2cm]{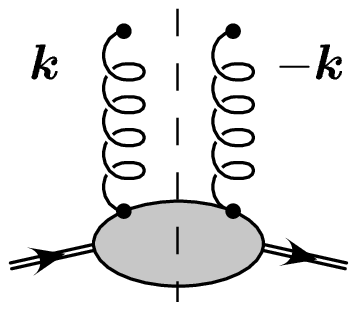}}
\end{align}
and take Fourier--transforms of each term in~\eqref{eq:N-low-density} with
respect to the relative momentum $k$ according to
\begin{align}
  \label{eq:one-term-FT}
  \parbox{2cm}{\includegraphics[width=2cm]{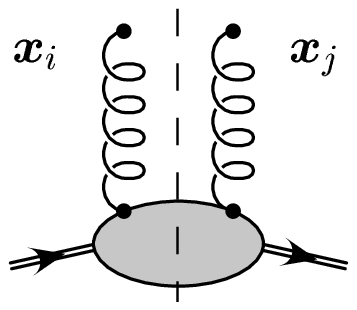}} =
  \int \frac{d^2k}{(2\pi)^2} \ e^{i \bm k \cdot\left({\bm x}_i-{\bm x}_j\right)}  
  \parbox{2cm}{\includegraphics[width=2cm]{Gdef-mom-cut-forward}}
\end{align}
The first two diagrams in~\eqref{eq:N-low-density}, with their degenerate
Fourier factors $e^{i{\bm k} \cdot {\bm x}_{00}}$ and $e^{i{\bm k} \cdot {\bm
    x}_{11}}$ contribute the $+1$ term. For the second term we also use that
$N(k,Y)$ only depends on $k=|\bm k|$, following simplifying assumption made
at the outset of this section.

Substituting \eq{FT} into \eq{BFKL} yields
\begin{align}\label{BFKL2}
  \int \frac{d^2 k}{(2 \pi)^2} \, (1 - e^{i {\bm k} \cdot {\bm
      x}_{01}}) \, \frac{\partial {\tilde N} (k,Y)}{\partial Y} \, =
  \, \frac{\as \, N_c}{2 \, \pi^2} \, \int \frac{d^2 k}{(2 \pi)^2} \,
  {\tilde N} (k,Y) \notag \\ \times \, \int d^2 x_2 \,
  \frac{x_{01}^2}{x_{20}^2 \, x_{21}^2} \, \left[1 - e^{- i {\bm k}
      \cdot {\bm x}_{02}} - e^{- i {\bm k} \cdot {\bm x}_{12}} + e^{i
      {\bm k} \cdot {\bm x}_{01}} \right].
\end{align}
To obtain a momentum space expression we now use~(\ref{BFKL2}), along
with the Fourier--representation of the dipole kernel
\begin{equation}\label{dipm}
  \alpha_s\frac{x_{01}^2}{x_{20}^2 \, x_{21}^2} \, = \, \int \frac{d^2
    q \, d^2 q'}{(2 \pi)^2} K_{{\bm q}, {\bm q}'} \, (e^{i {\bm q}
    \cdot {\bm x}_{02}} - e^{i {\bm q} \cdot {\bm x}_{12}}) \, (e^{- i
    {\bm q}' \cdot {\bm x}_{02}} - e^{- i {\bm q}' \cdot {\bm
      x}_{12}}) \ , \hspace{1cm} K_{{\bm q}, {\bm q}'} := \alpha_s
  \frac{ {\bm q} \cdot {\bm q}' }{ {\bm q}^2 \, {\bm q}'^2 }\ ,
\end{equation}
and insert both into the low density expanded BK equation~(\ref{BFKL}) to
obtain
\begin{align}
  \label{eq:insertionstep-strategy}
  \int\!\frac{d^2k}{(2\pi)^2} & \left(1 - e^{i{\bm k}\cdot{\bm
        x}_{01}} \right) \, \frac{\partial \Tilde N(k,Y)}{\partial Y}
  = \frac{ N_c}{2\pi^2} \int d^2{\bm x}_2 \int \frac{d^2 {q}\, d^2
    {{q}}'}{(2\pi)^2} \frac{d^2k}{(2\pi)^2} \ K_{\bm{q}, {\bm{q}}'}
  \notag \\ & \times \left(e^{i \bm{q}{\cdot}{\bm{x}}_0}-e^{i
      \bm{q}{\cdot}{\bm{x}}_1}\right) \left( e^{-i {\bm{q}}'
      {\cdot}{\bm{x}}_1} -e^{-i {\bm{q}}' {\cdot}{\bm{x}}_0} \right) \ 
  e^{-i({\bm q}-{\bm q}')\cdot{\bm x}_2} \notag \\ & \times \left[
    \left(1 - e^{i{\bm k}\cdot {\bm x}_{02}}\right) +\left(1 -
      e^{-i{\bm k}\cdot {\bm y}_{12}} \right) -\left(1 - e^{i{\bm
          k}\cdot {\bm x}_{01}} \right) \right] \, \Tilde N(k,Y) \ .
\end{align}
The goal now is to write the right hand side in a form that resembles the
structure of the left hand side:
\begin{equation*}
  \int\!\frac{d^2k}{(2\pi)^2}  
  \left(1 - e^{i{\bm k}\cdot {\bm x}_{01}} \right)\ 
  \left( \text{kernels and $N$ dependence of r.h.s.} \right)
\ .
\end{equation*}
Once this is done, one may read off the momentum space equation for
$\Tilde N(k,Y)$ by equating the integrands.

To do so one first integrates over ${\bm x}_2$ to eliminate ${\bm{q}}'$
and then transforms the remaining contributions (by shifting ${\bm{q}}$ and
reflecting its sign as needed on a term by term basis) until the result for
the right hand side takes the form
\begin{equation*}
\frac{N_c}{2\pi^2}\int d^2q\, d^2k\ \left\{
    \left(1 - e^{i {\bm q}\cdot {\bm x}_{01} }\right)
         K_{\text{real}}({\bm{q}},{\bm{k}}) \, \Tilde N({\bm{k}},Y)
    + \left(1 - e^{i {\bm k}\cdot {\bm x}_{01}} \right)
         K_{\text{virt}}({\bm{k}},{\bm{q}}) \, \Tilde N({\bm{k}},Y)
        \right\}.
\end{equation*}
Last, we exchange integration variables ${\bm k}\leftrightarrow{\bm q}$
in the real term and equate the integrands of the resulting equation:
\begin{align}
  \label{eq:abstract-BFKL-1}
  \frac{\partial \Tilde N({\bm k},Y)}{\partial Y} \, = \,
  \frac{N_c}{2\pi^2} \int d^2q\ \left\{
    K_{\text{real}}({\bm{k}},{\bm{q}}) \, \Tilde N({\bm{q}},Y) +
    K_{\text{virt}}({\bm{k}},{\bm{q}}) \, \Tilde N({\bm{k}},Y)
  \right\} \ .
\end{align}
{\em Without} using any specific knowledge about the form of the kernel
$K_{{\bm q},\Tilde{\bm q}}$ we find
that the Fourier-kernels $K_{{\bm q},\Tilde{\bm
    q}}$ only enter in combinations
\begin{align}
  \label{eq:Ldef}
  L_{{\bm q},\Tilde{\bm q}} := K_{{\bm q},{\bm q}}-K_{{\bm q},\Tilde{\bm q}}
\end{align}
that vanish in the limit ${{\bm q}\to\Tilde{\bm q}}$.
In terms of $L_{{\bm q},\Tilde{\bm q}}$ the kernels take the form
\begin{subequations}
  \label{eq:G-kernels}
\begin{align}
  K_{\text{real}}({\bm{k}},{\bm{q}}) = & L_{\bm{q}-{\bm k},-{\bm
      k}}+L_{-{\bm k},\bm{q}-{\bm k}} +L_{{\bm k},\bm{k}-{\bm
      q}}+L_{{\bm k}-{\bm q},{\bm k}}
  \\
  K_{\text{virt}}({\bm{k}},{\bm{q}}) = & L_{{\bm q},\bm{q}-{\bm
      k}}+L_{{\bm q},\bm{q}+{\bm k}}.
\end{align}
\end{subequations}
We will exploit this generic structure below when we consider running coupling
corrections.  For now, substituting the leading order form for $K_{{\bm
    q},\Tilde{\bm q}}$ into~(\ref{eq:Ldef}) yields:
\begin{align}
  L_{{\bm q},\Tilde{\bm q}} = \frac{\alpha_s}2\left\{
\frac{1}{{\bm q}^2}
-\frac{1}{\Tilde{\bm q}^2}
+\frac{({\bm q}-\Tilde{\bm q})^2}{{\bm q}^2 \Tilde{\bm q}^2}
\right\}
\ .
\end{align}
This then is used in~(\ref{eq:Ldef}) and~(\ref{eq:abstract-BFKL-1}).  It is
worth noting that only the part symmetric under exchange of ${\bm q}$ and
$\Tilde{\bm q}$ contributes to~(\ref{eq:abstract-BFKL-1}), the other
contributions cancel out. We find
\begin{align}\label{BFKL6}
  \frac{\partial {\tilde N} (k,Y)}{\partial Y} \, = \, \frac{\as \,
    N_c}{2 \, \pi^2} \, \int d^2 q \, \left[ \frac{2 \, {\bm
        q}^2}{{\bm k}^2 \, ({\bm k} - {\bm q})^2} \, {\tilde N} (q, Y)
    - \frac{{\bm k}^2}{{\bm q}^2 \, ({\bm k} - {\bm q})^2} \, {\tilde
      N} (k, Y) \right].
\end{align}

Finally, defining the unintegrated gluon distribution via
\begin{align}\label{phi}
  \as (k^2) \, \phi (k,Y) \, = \, \frac{N_c \, S_\perp}{(2 \pi)^3} \, k^2 \,
  {\tilde N} (k,Y)
\end{align}
we write
\begin{align}\label{BFKL7}
  \frac{\partial \phi (k,Y)}{\partial Y} \, = \, \frac{\as \, N_c}{2
    \, \pi^2} \, \int d^2 q \, \left[ \frac{2}{({\bm k} - {\bm q})^2}
    \, \phi (q, Y) - \frac{{\bm k}^2}{{\bm q}^2 \, ({\bm k} - {\bm
        q})^2} \, \phi (k, Y) \right].
\end{align} 
At the leading logarithmic order we have neglected the running of
$\as$ and dropped this factor on both sides of \eq{BFKL7}.  \eq{BFKL7}
is the momentum space representation of the dipole model analogue of
the BFKL equation.

The definition of the unintegrated gluon distribution $\phi (k,Y)$
given by Eqs. (\ref{phi}) and (\ref{FT}) is illustrated by the figure
of \eq{eq:N-low-density}.  Since we are interested in the linear (low
color density) regime the interaction between the dipole and the
target is limited to the exchange of a single BFKL (or, as we will
consider later, NLO BFKL) ladder. Thus the unintegrated gluon
distribution $\phi (k,Y)$ includes the whole (NLO) BFKL ladder
attached to the target. Summation over all possible connections of the
$t$-channel gluons to the quark and anti-quark lines in the dipole has
to be performed, as shown in \eq{eq:N-low-density}. Due to our
assumption of independence of the dipole amplitude $N$ on the impact
parameter shown in \eq{bindep} we were able to explicitly integrate
over the impact parameter in \eq{phi} which resulted in the factor of
the transverse area of the target $S_\perp$.

The coupling constant $\as (k^2)$ accounts for the interaction of the
$t$-channel gluons with the quark lines of the dipole: indeed this
interaction is not a part of the unintegrated gluon distribution
function and has to be factored out. The scale of the running coupling
is naturally given by ${\bm k}^2$ as the only available scale in the
problem. As we will see later, even at the NLO BFKL level one does not
need to know the constant under the logarithm in this factor of the
running coupling.  If one takes the diagrams in \eq{eq:N-low-density}
literally and tries to calculate the contribution of the two exchanged
gluons with the running coupling corrections included, one would get a
factor of $(\as ({\bm k}^2)^2 /\am ) \, 1/{\bm k}^4$, with $\am$ the bare
coupling constant.  Here we follow the standard convention for the
unintegrated gluon distribution function and require that at the
lowest (two gluon) level it should be given by
\begin{align}\label{phi_lo}
\phi_0 (k,Y) \, = \, \frac{\as \, C_F}{\pi} \, \frac{1}{{\bm k}^2}.
\end{align}
Unintegrated gluon distribution defined by \eq{phi_lo} actually gives
us the number of gluon quanta in the phase space region specified by
its arguments.  To adhere to this lowest order definition of the
unintegrated gluon distribution we absorb the factor of $(\as (k^2)
/\am ) \, 1/{\bm k}^2$ coming from the two gluons in the figure of
\eq{eq:N-low-density} into $\phi (k,Y)$, leaving the factor of $\as
(k^2) / {\bm k}^2$ out in front, as we see in \eq{phi}.


\subsection{Leading order intercept}
\label{LOint}

\eq{BFKL7} is equivalent to the BFKL equation \cite{Bal-Lip,Kuraev:1977fs},
averaged over azimuthal angles. To see that this is indeed the case, and to
prepare for the pomeron intercept calculations to be carried out below, let us
show that powers of momentum are indeed the eigenfunctions of the kernel
of~\eq{BFKL7} and find the corresponding eigenvalues. Acting on $\phi (k, Y)
\sim k^{2 \, \lambda}$ with the kernel of \eq{BFKL7} yields
\begin{align}
  \frac{\as \, N_c}{\pi^2} \, \int d^2 q \, K^{\text{LO}} ({\bm k},
  {\bm q}) \, q^{2 \, \lambda} \, \equiv \, \frac{\as \, N_c}{2 \,
    \pi^2} \, \int d^2 q \, \left[ \frac{2}{({\bm k} - {\bm q})^2} \,
    q^{2 \, \lambda} - \frac{{\bm k}^2}{{\bm q}^2 \, ({\bm k} - {\bm
        q})^2} \, k^{2 \, \lambda} \right].
\end{align}
Rewriting the measure of the $\bm q$-integral as \cite{Mueller:1994rr}
\begin{align}\label{measure}
  d^2 q \, = \, 2 \pi dq \, q \, dl \, l \, \int\limits_0^\infty db \,
  b \, J_0 (b \, k) \, J_0 (b \, q) \, J_0 (b \, l)
\end{align}
with $k = |{\bm k}|$, $q = |{\bm q}|$, $l = |{\bm k} - {\bm q}|$ we
obtain
\begin{align}\label{loeig1}
  \int d^2 q \, K^{\text{LO}} ({\bm k}, {\bm q}) \, q^{2 \, \lambda}
  \, = \, 2 \, \pi \, \int\limits_0^\infty db \, b \, J_0
  (b \, k) \, \int\limits_\Lambda^\infty dq \, q \, J_0 (b \, q) \,
  \int\limits_\Lambda^\infty dl \, l^{-1} \, J_0 (b \, l) \, \left[
    q^{2 \, \lambda} - \frac{1}{2 \, {q}^2} \, k^{2 \, \lambda + 2}
  \right],
\end{align}
where we introduced an infrared cutoff $\Lambda$ to regulate $q$- and
$l$-integrals. Using Eqs. (\ref{master}) and (\ref{B2}) in Appendix
\ref{App2} we can perform the $q$- and $l$-integrals recasting
\eq{loeig1} into
\begin{align}\label{loeig2}
  & \int d^2 q \, K^{\text{LO}} ({\bm k}, {\bm q}) \, q^{2 \, \lambda}
  \notag \\ & = \, 2 \, \int\limits_0^\infty db \, b \, J_0 (b \, k)
  \, \left( \psi (1) - \ln \frac{b \, \Lambda}{2} \right) \, \left[
    2^{2 \, \lambda + 1} \, b^{- 2 \, \lambda -2} \, \frac{\Gamma
      \left( 1 + \lambda \right)}{\Gamma \left( - \lambda \right)} -
    \frac{k^{2 \, \lambda + 2}}{2} \, \left( \psi (1) - \ln \frac{b \,
        \Lambda}{2} \right) \right].
\end{align}
After performing the integration over $b$ (again with the help of the
formulae from Appendix \ref{App2}) \eq{loeig2} reduces to
\begin{align}\label{loeig3}
  \frac{\as \, N_c}{\pi} \, \int d^2 q \, K^{\text{LO}} ({\bm k}, {\bm
    q}) \, q^{2 \, \lambda} \, = \, \frac{\as \, N_c}{\pi} \, \chi
  (-\lambda) \, k^{2 \, \lambda}
\end{align}
where
\begin{align}\label{chi}
  \chi (-\lambda) \, = \, 2 \, \psi (1) - \psi (-\lambda) - \psi (1 +
  \lambda).
\end{align}
This accomplishes the proof that, at least in the azimuthally
symmetric case, \eq{BFKL7} is equivalent to the BFKL equation.


\section{Momentum space NLO BFKL equation with \\ 
the leading-$N_f$ NLO correction in the kernel}
\label{momNLO}

\begin{figure}[htb]
  \centering
  \begin{minipage}{5.2cm}  
\centering
  \includegraphics[width=5.2cm]{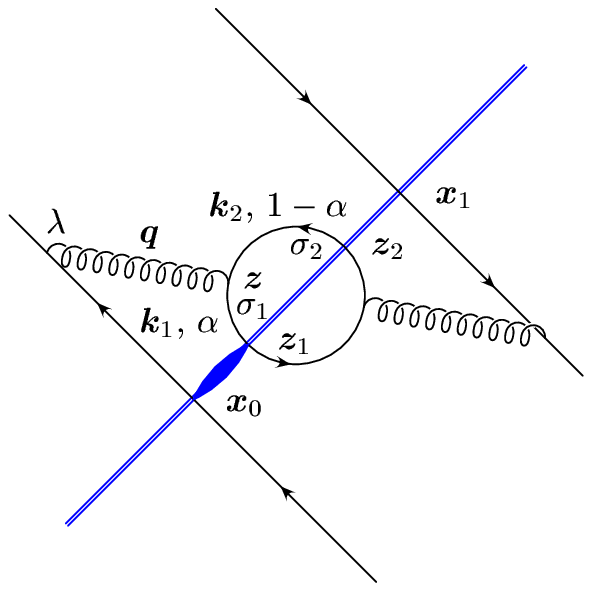}
\\ A
\end{minipage}  
\begin{minipage}{5.2cm} 
\centering 
\includegraphics[width=5.2cm]{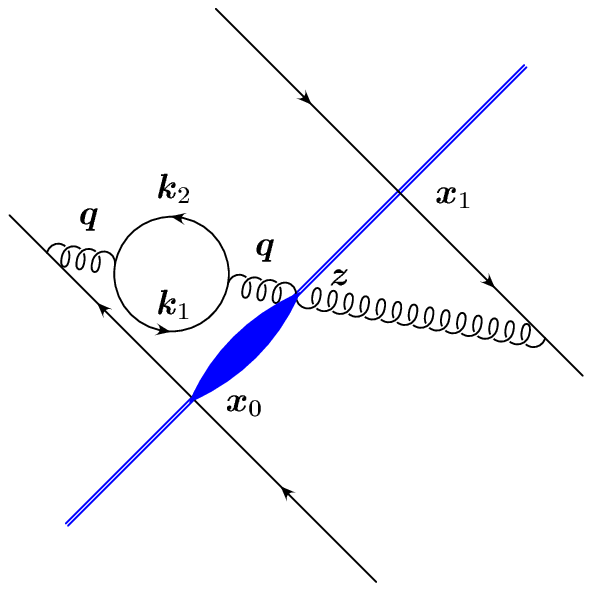}
\\ B
\end{minipage} 
    \begin{minipage}{5.2cm}
\centering
\includegraphics[width=5.2cm]{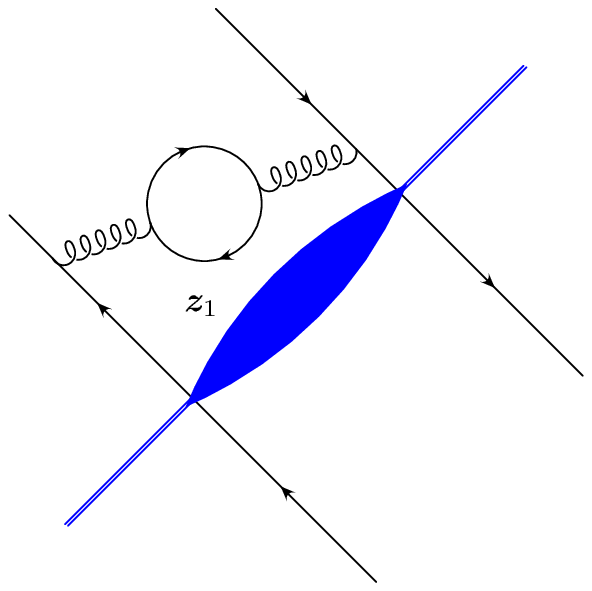}
\\ C
  \end{minipage}

  \caption{Diagrams contributing a factor $N_f$ at order $\alpha_s^2
    \ln(1/x)$.  The line from bottom left to top right indicates the
    target field crossed by the fast moving constituents of the
    projectile. Their transverse positions are fixed. The ovals
    indicate insertions of $N({\bm x}_0,{\bm z}_1,Y)$, $N({\bm
      x}_0,{\bm z},Y)$ and $N({\bm x}_0,{\bm x}_1,Y)$ respectively.}
  \label{fig:gendiags}
\end{figure}

To calculate the leading-$N_f$ contribution to the NLO BFKL kernel in the
language of Mueller's dipole model
\cite{Mueller:1994rr,Mueller:1994jq,Mueller:1995gb} we have to insert a single
quark bubble correction in the gluon line emitted in one step of the dipole
evolution. (For a brief review of Mueller's dipole model see
\cite{Jalilian-Marian:2005jf}.) The relevant diagrams are shown in
Fig.~\ref{fig:gendiags} and have been calculated recently by the authors in
\cite{HW-YK:2006} and by Balitsky in \cite{Bal:2006}.\footnote{We emphasize
  that the derivation there was performed in the JIMWLK/BK context that
  includes nonlinear contributions. Here we only retain the linear terms
  relevant for the BFKL limit. Using light-cone perturbation theory one might
  have calculated these contributions directly in the dipole model.}

The calculation in \cite{HW-YK:2006} followed the rules of the light cone
perturbation theory (LCPT) \cite{Lepage:1980fj,Brodsky:1997de}. The diagrams
in Fig.~\ref{fig:gendiags} should also be understood as LCPT diagrams. The
propagators of gluon lines in the graphs A and C in Fig.~\ref{fig:gendiags}
also include the instantaneous parts.

Here we are interested in the linear evolution: hence only one of the
produced dipoles in Figs.~\ref{fig:gendiags} A and B continues to have
evolution corrections and interacts with the target. In
Fig.~\ref{fig:gendiags} we show that with an oval denoting the
interacting dipole. Indeed either one of the dipoles can interact and
later on we will add the contribution with the other dipole
interacting with the target.

We start with the contribution to the NLO JIMWLK kernel coming from
the diagram A in Fig.~\ref{fig:gendiags}, which is given by Eq.~(17)
in \cite{HW-YK:2006}:
\begin{align}\label{K1JIMWLKmom}
  {\cal K}_1^{\text{NLO}} ({\bm x}_0, {\bm x}_1 ; {\bm z}_1, {\bm
    z}_2) \, = & \, 4 \, N_f \, \int\limits_0^1 d \alpha \, \int
  \frac{d^2 k}{(2\pi)^2}\frac{d^2 k'}{(2\pi)^2}\frac{d^2 q}{(2\pi)^2}
  \frac{d^2 q'}{(2\pi)^2} \ e^{ -i {\bm q}\cdot ({\bm z}-{\bm x}_0) +i
    {\bm q}' \cdot ({\bm z}-{\bm x}_1) -i({\bm k}-{\bm k}') \cdot {\bm
      z}_{12}} 
\notag \\ & \displaybreak[0]\times \left[ \frac{1}{{\bm q}^2{\bm
        q}^{\prime 2}} \frac{(1-2\alpha)^2 {\bm q}\cdot{\bm k}\ {\bm k}'
      \cdot {\bm q}' + {\bm q} \cdot {\bm q}' \ {\bm k}\cdot{\bm k}' -
      {\bm q}\cdot{\bm k}' \ {\bm k}\cdot{\bm q}'}{\Big[{\bm k}^2+{\bm
        q}^2\alpha(1-\alpha)\Big]\Big[{\bm k}'^2+{\bm
        q}'^2\alpha(1-\alpha)\Big]} \right. 
\notag \\ &\displaybreak[0] + \frac{2 \,
    \alpha \, (1-\alpha) \, (1-2 \alpha)}{\Big[{\bm k}^2+{\bm
      q}^2\alpha(1-\alpha)\Big]\Big[{\bm k}'^2+{\bm
      q}'^2\alpha(1-\alpha)\Big]} \, \left( \frac{{\bm k} \cdot {\bm
        q}}{{\bm q}^2} + \frac{{\bm k}' \cdot {\bm q}'}{{\bm q}'^2}
  \right)\notag \\ & \left. + \, \frac{4 \, \alpha^2 \,
      (1-\alpha)^2}{\Big[{\bm k}^2+{\bm
        q}^2\alpha(1-\alpha)\Big]\Big[{\bm k}'^2+{\bm
        q}'^2\alpha(1-\alpha)\Big]} \right].
\end{align}
At the moment we exclude all the factors of the coupling constant and
the color factor, similar to what was done in~\cite{HW-YK:2006}.
The notations used in \eq{K1JIMWLKmom} are explained in
\fig{fig:gendiags}A: the quark and the anti-quark have transverse
momenta ${\bm k}_1$ and ${\bm k}_2$ and transverse coordinates ${\bm
  z}_1$ and ${\bm z}_2$.  The gluon line carries momentum ${\bm q}$ in
the amplitude. All the transverse momenta in the complex conjugate
amplitude are labeled in the same way as momenta in the amplitude only
with a prime.  The longitudinal fraction of the gluons' momentum
carried by the quark is labeled $\alpha = k_{1+} / (k_{1+} + k_{2+})$.
We also note that ${\bm z} = \alpha \, {\bm z}_1+(1-\alpha) \, {\bm
  z}_2$, ${\bm z}_{12} = {\bm z}_1 -{\bm z}_2$, ${\bm q} = {\bm
  k}_1+{\bm k}_2$ and ${\bm k}={\bm k}_1(1-\alpha)-{\bm k}_2\,
\alpha$.

Using ${\bm q} = {\bm k}_1+{\bm k}_2$ and ${\bm k}={\bm
  k}_1(1-\alpha)-{\bm k}_2\, \alpha$ (and the same for primed ones) we
rewrite \eq{K1JIMWLKmom} as
\begin{align}\label{K1'}
  {\cal K}_1^{\text{NLO}} ({\bm x}_0, {\bm x}_1 ; {\bm z}_1, {\bm
    z}_2) \, = \, 4 \, N_f \, \int\limits_0^1 d \alpha \, \int
  \frac{d^2 k_1}{(2\pi)^2}\frac{d^2 k_2}{(2\pi)^2}\frac{d^2
    k'_1}{(2\pi)^2} \frac{d^2 k'_2}{(2\pi)^2} \notag \\ \times \, e^{
    -i {\bm k}_1 \cdot ({\bm z}_1 - {\bm x}_0) -i {\bm k}_2 \cdot
    ({\bm z}_2 - {\bm x}_0) +i {\bm k}'_1 \cdot ({\bm z}_1 - {\bm
      x}_1) + i {\bm k}'_2 \cdot ({\bm z}_2 - {\bm x}_1)} \,
  \frac{1}{({\bm k}_1 + {\bm k}_2)^2 \, ({\bm k}'_1 + {\bm k}'_2)^2} \notag \\
  \times \, \frac{{\bm k}_1 \cdot {\bm k}'_1 \, {\bm k}_2 \cdot {\bm
      k}'_2 - {\bm k}_1 \cdot {\bm k}'_2 \, {\bm k}'_1 \cdot {\bm k}_2
    + [{\bm k}_1 \cdot {\bm k}_2 + {\bm k}_1^2 \, (1-\alpha) + {\bm
      k}_2^2 \,\alpha] \, [{\bm k}'_1 \cdot {\bm k}'_2 + {\bm
      k}_1^{\prime 2} \, (1-\alpha) + {\bm k}_2^{\prime 2} \,\alpha]}{[{\bm k}_1^2
    \, (1-\alpha) + {\bm k}_2^2 \,\alpha] \, [{\bm k}_1^{\prime 2} \,
    (1-\alpha) + {\bm k}_2^{\prime 2} \,\alpha]}.
\end{align}

As indicated in \fig{fig:gendiags}A, in the linear regime only one of the
dipoles, say the dipole formed by ${\bm x}_0$ and ${\bm z}_1$ continues the
evolution.  Hence there is no dependence on the coordinate ${\bm z}_2$ in the
resulting $N(|{\bm x}_0 - {\bm z}_1|, Y)$. Therefore we can (and should)
integrate ${\bm z}_2$ out.  Adding the result to the other $N_f$ corrections
leads us to consider the whole contribution as part of the running coupling
corrections.

Let us note that this procedure is equivalent to the ``UV subtraction''
performed by Balitsky in~\cite{Bal:2006} for the nonlinear evolution equation
and briefly explain the context. For the nonlinear equations, extraction of
the UV-divergent part of the diagram A in~\fig{fig:gendiags}, which one
\emph{must} absorb in the running coupling constant, can be done in many ways,
depending on which linear combination of ${\bm z}_1$ and ${\bm z}_2$ is kept
fixed. This choice is referred to in~\cite{HW-YK:2006}, maybe somewhat
unfortunately, as the choice of ``subtraction point.'' Since the notions of
``UV subtraction'' and ``subtraction point'' play an important role in the
renormalization of QCD, we should attempt to forestall unnecessary confusion
by pointing out that the ideas involved here are only peripherally connected
to renormalization. The subtractions referred to here correspond to different
schemes of separating the running coupling contributions (which must carry the
UV divergence of QCD diagrams so that the UV scale $\mu$ will enter the
logarithms that induce the running of the coupling) from new physics channels
(such as, in this case, the presence of a well separated $\Bar q q$-pair in
the final state), from which the UV divergent running coupling contribution
has been subtracted. The only new channel we have considered explicitly, the
$\Bar q q$ channel, then turns out to be completely finite and $\mu$
independent at present accuracy.  This separation scheme dependence (and the
coordinate space ``subtraction points'' associated with it) is a freedom that
arises independently {\em in addition} to the renormalization scheme
dependence and the momentum space renormalization scale $\mu$ (with associated
subtraction points) appearing in the renormalization procedure of QCD. In
the case of non-linear evolution the arbitrariness of the separation scheme
results in the differences between~\cite{HW-YK:2006} and~\cite{Bal:2006}. We
emphasize that the \emph{total} result, in which running coupling
contributions and new channels are added together, is \emph{not} affected by
this separation. In contrast to renormalization scheme dependence, which can
not be eliminated at any finite order in perturbation theory, this statement
about separation scheme independence holds order by order in perturbation
theory. In the separation of~\cite{HW-YK:2006} both the running coupling
corrections and the new $q\Bar q$ channel contribute leading-$N_f$ terms to
the linear part of the BK equation and the sum of these contributions is
obtained by the the above procedure. In the separation of~\cite{Bal:2006}, all
leading order $N_f$--contributions to the linear part of the BK equation are
included into the running coupling, while the correspondingly constructed new $q\Bar
q$ channel only contributes a nonlinear term to the BK equation. This establishes the
direct correspondence of our present procedure (which is simply aimed at
collecting all leading-$N_f$ contributions to the linear part of the BK
equation) to the subtraction used in~\cite{Bal:2006}.

Let us stress again that for the total leading-$N_f$ contribution (and that
is what we are calculating here) subtraction points are not an issue, they
merely affect the separation into what we interpret as a running coupling
correction and new production channels. The only remarkable point here is that
the leading-$N_f$ contributions may \emph{all} be interpreted as running
coupling corrections.

To return to our present task: we are interested in the linear evolution and
have to integrate the kernel in~\eq{K1'} either over ${\bm z}_2$ or ${\bm
  z}_1$ depending on which dipole interacts with the target.

To integrate over ${\bm z}_2$, which would put ${\bm k}'_2 = {\bm
  k}_2$, we define ${\bm q} = {\bm k}_1+{\bm k}_2$ and ${\bm q}' =
{\bm k}'_1+{\bm k}'_2 = {\bm k}'_1+{\bm k}_2$, obtaining
\begin{align}\label{K1Bal}
  \int d^2 z_2 \, {\cal K}_1^{\text{NLO}} ({\bm x}_0, {\bm x}_1 ; {\bm
    z}_1, {\bm z}_2) \, = \, 4 \, N_f \, \int\limits_0^1 d \alpha \,
  \int \frac{d^2 q}{(2\pi)^2}\frac{d^2 q'}{(2\pi)^2}\frac{d^2
    k}{(2\pi)^2} \ e^{ -i {\bm q} \cdot ({\bm z}_1 - {\bm x}_0) + i
    {\bm q} \cdot ({\bm z}_1 - {\bm x}_1)} \notag \\
  \times \, \frac{1}{{\bm q}^2 \, {\bm q}^{\prime 2} \, [({\bm q}- {\bm
      k})^2 \, (1-\alpha) + {\bm k}^2 \,\alpha] \, [ ({\bm q}' - {\bm
      k})^2 \, (1-\alpha) + {\bm k}^2 \,\alpha]} \notag \\ \times \,
  \bigg\{ {\bm k}^2 \, ({\bm q}- {\bm k}) \cdot ({\bm q}' - {\bm k}) -
  {\bm k} \cdot ({\bm q} - {\bm k}) \, {\bm k} \cdot ({\bm q}' - {\bm
    k}) + [{\bm k} \cdot ({\bm q} - {\bm k}) + ({\bm q}- {\bm k})^2 \,
  (1-\alpha) + {\bm k}^2 \,\alpha] \notag \\ \times \, [ {\bm k} \cdot
  ({\bm q}' - {\bm k}) + ({\bm q}' - {\bm k})^2 \, (1-\alpha) + {\bm
    k}^2 \,\alpha] \bigg\}
\end{align}
where we dropped the subscript $2$ in ${\bm k}_2$, i.e., ${\bm k}_2
\rightarrow {\bm k}$. Now we have to integrate over ${\bm k}$. To do
so we first write
\begin{align}\label{K1Bal2}
  \int d^2 z_2 \, {\cal K}_1^{\text{NLO}} ({\bm x}_0, {\bm x}_1 ; {\bm
    z}_1, {\bm z}_2) \, = \, 4 \, N_f \, \int\limits_0^1 d \alpha \,
  \int \frac{d^2 q}{(2\pi)^2}\frac{d^2 q'}{(2\pi)^2}\frac{d^d
    k}{(2\pi)^d} \ e^{ -i {\bm q} \cdot ({\bm z}_1 - {\bm x}_0) + i
    {\bm q} \cdot ({\bm z}_1 - {\bm x}_1)} 
\notag \\ \displaybreak[0]
  \times \, \frac{1}{{\bm q}^2 \, {\bm q}^{\prime 2}} \, \Bigg\{ \frac{{\bm
      k}^2 \, ({\bm q}- {\bm k}) \cdot ({\bm q}' - {\bm k})}{[({\bm
      q}- {\bm k})^2 \, (1-\alpha) + {\bm k}^2 \,\alpha] \, [ ({\bm
      q}' - {\bm k})^2 \, (1-\alpha) + {\bm k}^2 \,\alpha]} \notag \\
  + \frac{{\bm k} \cdot ({\bm q} - {\bm k})}{({\bm q}- {\bm k})^2 \,
    (1-\alpha) + {\bm k}^2 \,\alpha} + \frac{{\bm k} \cdot ({\bm q}' -
    {\bm k})}{({\bm q}' - {\bm k})^2 \, (1-\alpha) + {\bm k}^2 \,
    \alpha} +1 \Bigg\},
\end{align}
where, in anticipation of dimensional regularization we have inserted
the dimension $d$ of the ${\bm k}$-integral explicitly.  The details
of ${\bm k}$-integration in \eq{K1Bal2} are given in Appendix
\ref{App1}. The result yields
\begin{align}\label{K1Bal3}
  \int d^2 z_2 \, {\cal K}_1^{\text{NLO}} ({\bm x}_0, {\bm x}_1 ; {\bm
    z}_1, {\bm z}_2) \, = \, \frac{N_f}{3 \, \pi} \, \int \frac{d^2
    q}{(2\pi)^2}\frac{d^2 q'}{(2\pi)^2} \, \ e^{ -i {\bm q} \cdot
    ({\bm z}_1 - {\bm x}_0) + i {\bm q} \cdot ({\bm z}_1 - {\bm x}_1)}
  \notag \\ \times \, \frac{1}{{\bm q}^2 \, {\bm q}^{\prime 2}} \Bigg\{
  ({\bm q} - {\bm q}')^2 \, \ln \frac{({\bm q} - {\bm q}')^2 \,
    e^{-5/3}}{\mu_{\overline{\text{MS}}}^2} - {\bm q}^2 \, \ln
  \frac{{\bm q}^2 \, e^{-5/3}}{\mu_{\overline{\text{MS}}}^2} - {\bm
    q}^{\prime 2} \, \ln \frac{{\bm q}^{\prime 2} \,
    e^{-5/3}}{\mu_{\overline{\text{MS}}}^2} \Bigg\}.
\end{align}
Note that due to the symmetry property
\begin{align}
  {\cal K}_1^{\text{NLO}} ({\bm x}_0, {\bm x}_1 ; {\bm
    z}_1, {\bm z}_2) \, = \, {\cal K}_1^{\text{NLO}} ({\bm x}_0, {\bm x}_1 ; {\bm
    z}_2, {\bm z}_1)
\end{align}
the contribution to the NLO BFKL kernel of \fig{fig:gendiags}A with
the dipole formed by ${\bm z}_2$ and ${\bm x}_1$ interacting is equal
to the right hand side of \eq{K1Bal3} with ${\bm z}_1 \rightarrow {\bm
  z}_2$.

To construct the full leading-$N_f$ NLO BFKL kernel one has to add to
\eq{K1Bal3} the contribution to the NLO BFKL kernel ${\cal
  K}_2^{\text{NLO}}$ of the diagram B from \fig{fig:gendiags}. Using
Eq.  (28) in \cite{HW-YK:2006} along with its mirror reflection with
respect to the line denoting the interaction with the target we obtain
\begin{align}\label{NLOk1}
  \int d^2 z_2 \, {\cal K}_1^{\text{NLO}} ({\bm x}_0, {\bm x}_1 ; {\bm
    z}_1, {\bm z}_2) + {\cal K}_2^{\text{NLO}} ({\bm x}_0, {\bm x}_1 ;
  {\bm z}_1) \, = \, \frac{N_f}{3 \, \pi} \, \int \frac{d^2
    q}{(2\pi)^2}\frac{d^2 q'}{(2\pi)^2} \, \ e^{ -i {\bm q} \cdot
    ({\bm z}_1 - {\bm x}_0) + i {\bm q} \cdot ({\bm z}_1 - {\bm x}_1)}
  \notag \\ \times \, \frac{1}{{\bm q}^2 \, {\bm q}^{\prime 2}} \Bigg\{
  ({\bm q} - {\bm q}')^2 \, \ln \frac{({\bm q} - {\bm q}')^2 \,
    e^{-5/3}}{\mu_{\overline{\text{MS}}}^2} - {\bm q}^2 \, \ln
  \frac{{\bm q}^2 \, e^{-5/3}}{\mu_{\overline{\text{MS}}}^2} - {\bm
    q}^{\prime 2} \, \ln \frac{{\bm q}^{\prime 2} \,
    e^{-5/3}}{\mu_{\overline{\text{MS}}}^2} \notag \\ + 2 \, {\bm q}
  \cdot {\bm q}' \, \ln \frac{{\bm q}^2 \,
    e^{-5/3}}{\mu_{\overline{\text{MS}}}^2} + \, 2 \, {\bm q} \cdot
  {\bm q}' \, \ln \frac{{\bm q}^{\prime 2} \,
    e^{-5/3}}{\mu_{\overline{\text{MS}}}^2}\Bigg\}.
\end{align}
The above expression can be simplified to give
\begin{align}\label{NLOk2}
  \int d^2 z_2 \, {\cal K}_1^{\text{NLO}} ({\bm x}_0, {\bm x}_1 ; {\bm
    z}_1, {\bm z}_2) + {\cal K}_2^{\text{NLO}} ({\bm x}_0, {\bm x}_1 ;
  {\bm z}_1) \, = \, \frac{N_f}{3 \, \pi} \, \int \frac{d^2
    q}{(2\pi)^2}\frac{d^2 q'}{(2\pi)^2} \, \ e^{ -i {\bm q} \cdot
    ({\bm z}_1 - {\bm x}_0) + i {\bm q} \cdot ({\bm z}_1 - {\bm x}_1)}
  \notag \\ \times \, \frac{1}{{\bm q}^2 \, {\bm q}^{\prime 2}} \Bigg\{ {\bm
    q}^2 \, \ln \frac{({\bm q} - {\bm q}')^2}{{\bm q}^2} + {\bm
    q}^{\prime 2} \, \ln \frac{({\bm q} - {\bm q}')^2}{{\bm q}^{\prime 2}} + \, 2
  \, {\bm q} \cdot {\bm q}' \, \ln \frac{{\bm q}^2 \, {\bm
      q}^{\prime 2}}{({\bm q} - {\bm q}')^2 \, \mu^2}\Bigg\},
\end{align}
where we have also defined $\mu^2 = \mu_{\overline MS}^2 \, e^{5/3}$. 

We do not need to calculate explicitly the diagram in
\fig{fig:gendiags} C due to the real-virtual cancellations which lead
to the identity
\begin{align}\label{rv}
  \int d^2 z_2 \, {\cal K}_1^{\text{NLO}} ({\bm x}_0, {\bm x}_1 ; {\bm
    z}_1, {\bm z}_2) + {\cal K}_2^{\text{NLO}} ({\bm x}_0, {\bm x}_1 ;
  {\bm z}_1) + {\cal K}_3^{\text{NLO}} ({\bm x}_0, {\bm x}_1 ;
  {\bm z}_1) \, = \, 0.
\end{align}
Here
\begin{align}
  \int d^2 z_1 \, {\cal K}_3^{\text{NLO}} ({\bm x}_0, {\bm x}_1 ;
  {\bm z}_1) \nonumber
\end{align}
is the contribution of \fig{fig:gendiags} C represented as the
integral over the transverse coordinate of the (anti-)quark.

The full leading-$N_f$ NLO dipole kernel is defined via
\begin{align}\label{NLOk3}
  K^{\text{NLO}}_f ({\bm x}_0, {\bm x}_1 ; {\bm z}_1) \, = \,
  \frac{N_c}{2} \, \sum_{m,n=0}^1 \, (-1)^{m+n} \left\{ \int d^2 z_2
    \, {\cal K}_1^{\text{NLO}} ({\bm x}_m, {\bm x}_n ; {\bm z}_1, {\bm
      z}_2) + {\cal K}_2^{\text{NLO}} ({\bm x}_m, {\bm x}_n ; {\bm
      z}_1) \right\}
\end{align}
where we have included the color factor $N_c /2$ in the large-$N_c$
approximation and have summed over all possible emissions of the
gluon in \fig{fig:gendiags} off the quark and the anti-quark lines in
the original dipole.

Including the contributions of either one of the dipoles in Figs.
\ref{fig:gendiags} A and B interacting with the target we derive that
the additive NLO contribution to the right hand side of \eq{BFKL} is
given by
\begin{align}\label{NLOk4}
  \am^2 \, \int d^2 x_2 \, K^{\text{NLO}}_f ({\bm x}_0, {\bm x}_1 ;
  {\bm x}_2) \, \left[ N ({\bm x}_{0}, {\bm x}_2, Y) + N ({\bm x}_{2},
    {\bm x}_1, Y) - N ({\bm x}_{0}, {\bm x}_1, Y) \right],
\end{align}
where $\am$ is the bare coupling constant. In the discussion of LO
BFKL the running of the coupling was negligible: hence we did not
distinguish the bare coupling from the physical coupling and labeled
them $\as$. At the NLO level running coupling corrections become
important and we will start distinguishing between the two. 

Combining Eqs.~(\ref{NLOk2}) and~(\ref{NLOk3}) yields an additive NLO
modification of the LO dipole kernel from Eq.~\eqref{dipm} of the form
\begin{align}
  K_{{\bm q},{\bm q}'} \, \to \, \am \frac{ {\bm q} \cdot {\bm q}'
    +\am \frac{N_f}{3\pi}\left( {\bm q}^2 \, \ln \frac{({\bm q} - {\bm
          q}')^2}{{\bm q}^2} + {\bm q}^{\prime 2} \, \ln \frac{({\bm
          q} - {\bm q}')^2}{{\bm q}^{\prime 2}} + \, 2 \, {\bm q}
      \cdot {\bm q}' \, \ln \frac{{\bm q}^2 \, {\bm q}^{\prime
          2}}{({\bm q} - {\bm q}')^2 \, \mu^2} \right) }{ {\bm q}^2 \,
    {\bm q}'^2 }.
\end{align}
Since our manipulations leading to the evolution
equation~\eqref{eq:abstract-BFKL-1} did not make use of any special
properties of the kernel, we only need to form linear combinations
according to~\eqref{eq:G-kernels} and~\eqref{eq:Ldef} to find the
evolution equation for the unintegrated gluon distribution function
defined in~\eq{phi}
\begin{align}\label{NLO_BFKL_1}
  \as (k^2) \, \frac{\partial \phi (k, Y)}{\partial Y} \, = \,
  \frac{\am \, N_c}{\pi^2} \, \int d^2 q \left[ \frac{\as (q^2)}{({\bm
        k} - {\bm q})^2} \, \phi (q, Y) - \as (k^2) \, \frac{{\bm k}
      \cdot ({\bm k} - {\bm q})}{{\bm q}^2 \, ({\bm k} - {\bm q})^2}
    \, \phi (k, Y) \right] + \frac{\am^2 \, N_c \, N_f}{12 \, \pi^3}
  \notag \\ \times \, \int d^2 q \, \left[ \frac{2 \, \as (q^2)
    }{({\bm k} - {\bm q})^2} \, \ln \left( \frac{{\bm k}^2 \, ({\bm k}
        - {\bm q})^2}{{\bm q}^2 \, \mu^2 } \right) \, \phi (q, Y) -
    \frac{\as (k^2) \, {\bm k}^2}{{\bm q}^2 \, ({\bm k} - {\bm q})^2}
    \, \ln \left( \frac{{\bm q}^2 \, ({\bm k} - {\bm q})^2}{{\bm k}^2
        \, \mu^2 } \right) \, \phi (k, Y) \right].
\end{align}
Here we could not neglect the running of the coupling and kept the
factors of the physical coupling $\as$ introduced in the definition of
the unintegrated gluon distribution in \eq{phi}. However, as we are
interested in the NLO correction only, we should divide both sides of
\eq{NLO_BFKL_1} by $\as (k^2)$ and expand the kernel on the right hand
side of the resulting equation up to the second order in $\am$ using
\begin{align}\label{expans}
  \frac{\as (q^2)}{\as (k^2)} \, \approx \, 1 + \am \, \beta_2 \, \ln
  \frac{k^2}{q^2} \, \rightarrow \, 1 - \frac{\am \, N_f}{6 \, \pi} \,
  \ln \frac{k^2}{q^2}.
\end{align}
Here
\begin{align}\label{beta2}
  \beta_2 \, = \, \frac{11 \, N_c - 2 \, N_f}{12 \, \pi}
\end{align}
and we replaced $\beta_2 \rightarrow - N_f / 6 \pi$ since we are
interested in leading-$N_f$ contribution only.

With the expansion of \eq{expans} we finally obtain 
\begin{align}\label{NLO_BFKL}
  \frac{\partial \phi (k, Y)}{\partial Y} \, = \, \frac{\am \, N_c}{2
    \, \pi^2} \, \int d^2 q \left[ \frac{2}{({\bm k} - {\bm q})^2} \,
    \phi (q, Y) - \frac{{\bm k}^2}{{\bm q}^2 \, ({\bm k} - {\bm q})^2}
    \, \phi (k, Y) \right]
  + \frac{\am^2 \, N_c \, N_f}{12 \, \pi^3} \notag \\
  \times \, \int d^2 q \, \left[ \frac{2}{({\bm k} - {\bm q})^2} \,
    \ln \left( \frac{({\bm k} - {\bm q})^2}{\mu^2 } \right) \, \phi
    (q, Y) - \frac{{\bm k}^2}{{\bm q}^2 \, ({\bm k} - {\bm q})^2} \,
    \ln \left( \frac{{\bm q}^2 \, ({\bm k} - {\bm q})^2}{{\bm k}^2 \,
        \mu^2 } \right) \, \phi (k, Y) \right].
\end{align}
\eq{NLO_BFKL} is the BFKL equation with the leading-$N_f$ NLO
correction calculated in the large-$N_c$ approximation.


\section{Resummation of bubble diagrams: triumvirate \\
couplings for forward BFKL}

\label{sec:resumm-bubble-diagr-forward}
\label{rcBFKL}

The leading-$N_f$ NLO BFKL kernel in \eq{NLO_BFKL} is remarkably
simple, appearing to be more compact than the similar kernel obtained
using conventional perturbation theory in
\cite{Fadin:1998py,Camici:1997ta}. To understand the origin of this
simplicity let us point out that, using the techniques developed in
\cite{HW-YK:2006} one can resum the $\am \, N_f$ corrections to all
orders, which correspond diagrammatically to inserting an infinite
chain of quark loops on the gluon line emitted in one step of
small-$x$ evolution in the $s$-channel approach. Replacing $N_f
\rightarrow - 6 \, \pi \, \beta_2$ and absorbing all the corrections
into the running coupling constant we obtain the JIMWLK kernel with
resummed running coupling corrections \footnote{We refer to this kernel as the JIMWLK kernel since it can be used to construct the JIMWLK evolution equation with the coordinate space subtraction point of \cite{Bal:2006}. Below we will use its Fourier transform to momentum space as the kernel of the BFKL equation with running coupling corrections.}
\begin{align}\label{Kall}
  \am \, {\cal K}_{\text{rc}} ({\bm x}_0, {\bm x}_1 ; {\bm z}) \, = \,
  4 \, \int \frac{d^2 q}{(2\pi)^2} \frac{d^2 q'}{(2\pi)^2} \ e^{ -i
    {\bm q}\cdot ({\bm z}-{\bm x}_0) +i {\bm q}' \cdot ({\bm z}-{\bm
      x}_1) } \, \frac{{\bm q} \cdot {\bm q}'}{{\bm q}^2 \, {\bm
      q}'^{2}} \, \frac{ \as \left({\bm q}^2 \, e^{-5/3} \right) \,
    \as \left({\bm q}'^2 \, e^{-5/3} \right)}{ \as \left(Q^2 \,
      e^{-5/3} \right)} \ ,
\end{align}
where
\begin{align}\label{Q}
  \ln \frac{Q^2 \, e^{-5/3}}{\mu_{\overline{{\text{MS}}}}^2} \, \equiv
  \, \frac{1}{2 \, {\bm q} \cdot {\bm q}'} \, \Bigg\{ {\bm q}^2 \, \ln
  \frac{{\bm q}^2 \, e^{-5/3}}{\mu_{\overline{\text{MS}}}^2} + {\bm
    q}^{\prime 2} \, \ln \frac{{\bm q}^{\prime 2} \,
    e^{-5/3}}{\mu_{\overline{\text{MS}}}^2} - ({\bm q} - {\bm q}')^2
  \, \ln \frac{({\bm q} - {\bm q}')^2 \,
    e^{-5/3}}{\mu_{\overline{\text{MS}}}^2} \Bigg\}.
\end{align}
This expression leads to a remarkably simple form for the building blocks of
the forward kernels for the angular averaged case: We find
\begin{subequations}
  \label{eq:Lnlo}
\begin{align}
  K_{{\bm q},{\bm q}'} \to &  K^{\text{NLO}}_{{\bm q},{\bm q}'}
= 
\frac12\left\{
\frac{\alpha_s({\bm q}^2\,e^{-5/3})}{{\bm q}^2}
+\frac{\alpha_s({\bm q}'^2\,e^{-5/3})}{{\bm q}'^2}
-\frac{\alpha_s({\bm q}^2\,e^{-5/3})\alpha_s({\bm q}'^2\,e^{-5/3})}{
  \alpha_s(({\bm q}-{\bm q}')^2\,e^{-5/3})}
  \frac{({\bm q}-{\bm q}')^2}{{\bm q}^2 {\bm q}'^2}
\right\},
\\
 L_{{\bm q},{\bm q}'} \to &  L^{\text{NLO}}_{{\bm q},{\bm q}'}
= 
\frac12\left\{
\frac{\alpha_s({\bm q}^2\,e^{-5/3})}{{\bm q}^2}
-\frac{\alpha_s({\bm q}'^2\,e^{-5/3})}{{\bm q}'^2}
+\frac{\alpha_s({\bm q}^2\,e^{-5/3})\alpha_s({\bm q}'^2\,e^{-5/3})}{
  \alpha_s(({\bm q}-{\bm q}')^2\,e^{-5/3})}
  \frac{({\bm q}-{\bm q}')^2}{{\bm q}^2 {\bm q}'^2}
\right\}.
\end{align}
\end{subequations}
Since the kernel in the real part is an explicitly symmetric combination of
$L$'s, one can immediately read off (dropping the NLO superscript for brevity)
\begin{align}
  \label{eq:ian-real-kernel}
L_{\bm{q}-{\bm k},-{\bm k}}+ & L_{-{\bm k},\bm{q}-{\bm k}}
+L_{{\bm k},\bm{k}-{\bm q}}+L_{{\bm k}-{\bm q},{\bm k}} =
  2\ \frac{\alpha_s({\bm k}^2\,e^{-5/3})\ 
    \alpha_s(({\bm q}-{\bm k})^2\,e^{-5/3})}{
    \alpha_s({\bm q}^2\,e^{-5/3})}
  \frac{{\bm q}^2}{{\bm k}^2 ({\bm q}-{\bm k})^2}
\ .
\end{align}
For the virtual contribution one notes that the antisymmetric parts of both
terms integrate to zero (by a shift in ${\bm q}$). We reverse the sign of
${\bm q}$ in the
remainder of the second term and are left with
\begin{align}
  \label{eq:ian-virt-kernel}
\int\!d^2q\Big(  L_{{\bm q},\bm{q}-{\bm k}}+L_{{\bm q},\bm{q}+{\bm k}}\Big)
= \int\!d^2q\ \frac{
  \alpha_s({\bm q}^2\,e^{-5/3})\alpha_s(({\bm q}-{\bm k})^2\,e^{-5/3})}{
    \alpha_s({\bm k}^2\,e^{-5/3})}
  \frac{{\bm k}^2}{{\bm q}^2 ({\bm q}-{\bm k})^2}.
\end{align}

Collecting all the contributions one obtains an equation for $\Tilde N(k,Y)$:
\begin{align}
  \label{eq:ian-G-eq}
  \frac{\partial \Tilde N(k,Y)}{\partial Y} \, = \, \frac{N_c}{2\pi^2}
  \int d^2q\ \Bigg\{ & 2\ \frac{\alpha_s({\bm k}^2\,e^{-5/3})
    \alpha_s(({\bm q}-{\bm k})^2\,e^{-5/3})}{ \alpha_s({\bm
      q}^2\,e^{-5/3})} \frac{{\bm q}^2}{{\bm k}^2 ({\bm q}-{\bm k})^2}
  \Tilde N(q,Y) \notag \\ & -\frac{ \alpha_s({\bm q}^2\,e^{-5/3})
    \alpha_s(({\bm q}-{\bm k})^2\,e^{-5/3})}{ \alpha_s({\bm
      k}^2\,e^{-5/3})} \frac{{\bm k}^2}{{\bm q}^2 ({\bm q}-{\bm k})^2}
  \Tilde N(k,Y) \Bigg\}.
\end{align}

Using \eq{phi} in \eq{eq:ian-G-eq} yields the following evolution
equation for the unintegrated gluon distribution including running
coupling corrections to all orders
\begin{align}\label{rc_BFKL}
  \frac{\partial \phi (k, Y)}{\partial Y} \, = \, \frac{N_c}{2 \,
    \pi^2} \, \int d^2 q \, \Bigg\{ & \frac{2}{({\bm k} - {\bm q})^2} \,
  \as \left( ({\bm k} - {\bm q})^2 \, e^{-5/3} \right) \, \phi (q, Y)
  \nonumber \\ & - \frac{{\bm k}^2}{{\bm q}^2 \, ({\bm k} - {\bm q})^2}
  \, \frac{ \as \left({\bm q}^2 \, e^{-5/3} \right) \, \as \left(
      ({\bm k} - {\bm q})^2 \, e^{-5/3} \right)}{ \as \left({\bm k}^2
      \, e^{-5/3} \right)} \, \phi (k, Y) \Bigg\}.
\end{align}
Expanding \eq{rc_BFKL} in powers of $\am$ to order $\am^2$ would yield
\eq{NLO_BFKL}.  Hence, from the standpoint of linear evolution, the
simplicity of the leading-$N_f$ NLO BFKL kernel obtained here is due
to the fact that this kernel consists only of running coupling
corrections.

\eq{rc_BFKL} should be compared with the evolution equation for the
unintegrated gluon distribution derived in~\cite{Levin:1994di} (c.f. Eqs. (22)
and (23) there). At first sight~\eq{rc_BFKL} appears to disagree with these
expressions. However, defining a new function
\begin{align}\label{tphi}
  {\tilde \phi} (k, Y) \, = \, \frac{\phi (k, Y)}{\as \left( {\bm k}^2
      \, e^{-5/3} \right)}
\end{align}
we recast \eq{rc_BFKL} into
\begin{align}\label{rc_BFKL_Genya}
  \frac{\partial {\tilde \phi} (k, Y)}{\partial Y} \, = \,
  \frac{N_c}{2 \, \pi^2} \, \int d^2 q & \ \frac{ \as \left({\bm q}^2 \,
      e^{-5/3} \right) \, \as \left( ({\bm k} - {\bm q})^2 \, e^{-5/3}
    \right)}{ \as \left({\bm k}^2 \, e^{-5/3} \right)} 
\nonumber \\ &
  \times \, \Bigg\{ \frac{2}{({\bm k} - {\bm q})^2} \, {\tilde \phi}
  (q, Y) - \frac{{\bm k}^2}{{\bm q}^2 \, ({\bm k} - {\bm q})^2} \,
  {\tilde \phi} (k, Y) \Bigg\}.
\end{align}
\eq{rc_BFKL_Genya} is in agreement with Eqs.~(22) and~(23)
of~\cite{Levin:1994di}. \eq{tphi} links the evolving quantity to the
unintegrated gluon distribution.

Eq.~\eqref{rc_BFKL_Genya} was originally obtained in~\cite{Levin:1994di} by
calculating running coupling corrections to the virtual contributions in the
non-forward case. This result was then generalized to include running coupling
corrections also to the real contribution by \emph{postulating} the validity
of the bootstrap equation beyond leading order~\cite{Levin:1994di,
  Braun:1994mw}.  To understand if this postulate can be justified from our
calculations, we need to explore also the non-forward case.


\section{Generalization to non-forward BFKL:  triumvirate\\ 
  couplings and bootstrap ideas}
\label{sec:gener-nonf-bfkl}
\label{nonfor}

Now that the general techniques are already familiar we relax our restrictions
to forward BFKL and angular averaged results. This will lead us to abandon our
initial assumption of impact parameter independence. Practically this amounts
to use a Fourier--representation of the lowest order expansions of $N({\bm
  x}_0,{\bm x}_1,Y)$ shown in Eq.~\eqref{eq:N-low-density} using two
independent momenta. We define
\begin{align}
  \label{eq:G-def}
  G_{{\bm x}_i,{\bm x}_j} 
  =\parbox{3cm}{\includegraphics[width=3cm]{Gdef-coord-cut}}
;
\hspace{1cm}
  G_{{\bm k},{\bm k}'} 
  = \parbox{3cm}{\includegraphics[width=3cm]{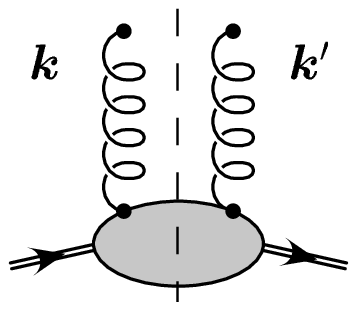}}
\ ,
\end{align}
which are related to each other by the double Fourier--transform
\begin{align}
  G_{{\bm x}_i,{\bm x}_j} = & \int\!\frac{d^2k}{(2\pi)^2} \frac{d^2k'}{(2\pi)^2}
  e^{i{\bm k}\cdot{\bm x}_i +i{\bm k}'\cdot{\bm x}_j}
  G_{{\bm k}, {\bm k}'}
=
\int\!\frac{d^2k}{(2\pi)^2} \frac{d^2l}{(2\pi)^2} 
  e^{i{\bm k}\cdot({\bm x}_i-{\bm x}_j) 
     +i{\bm l}\cdot\frac{{\bm x}_i+{\bm x}_j}2}
  G_{{\bm k}+\frac{{\bm l}}2, {\bm k}-\frac{{\bm l}}2 }
\ .
\end{align}
After the second equality sign we have changed integration variables to
conjugates of the relative and absolute positions in a form most suitable to
our calculations below. With this definition~\eq{FT} is replaced by
\begin{align}
\label{eq:nonforwar-N-via-G}
  N({\bm x}_i,{\bm x}_j,Y) = & \frac12
  \int\!\frac{d^2k}{(2\pi)^2} \frac{d^2l}{(2\pi)^2}  
  \left(
    e^{i {\bm l}\cdot{\bm x}_i}+e^{i {\bm l}{\bm x}_j}
    -e^{i{\bm k}\cdot{\bm x}_{i j}
      +i{\bm l}\cdot \frac{{\bm x}_i+{\bm x}_j}2}
    -e^{-i{\bm k}\cdot{\bm x}_{i j}
       +i{\bm l}\cdot \frac{{\bm x}_i+{\bm x}_j}2}
  \right) 
  G_{{\bm k}+\frac{{\bm l}}2, {\bm k}-\frac{{\bm l}}2 }(Y)
  \notag \\ &
  +\text{higher orders}
\end{align}
in a term by term correspondence with Eq.~\eqref{eq:N-low-density}.  Despite
this change compared to the forward case, one can still proceed to extract the
(non-forward) BFKL equation using a technique that completely parallels the
steps used in Sec.~\ref{DipBFKL}. The only price to pay is additional
algebraic effort. In this spirit, we insert our expressions for $N({\bm
  x}_i,{\bm x}_j,Y)$ and a generic kernel of the form~\eqref{dipm} into the
linearized BK equation~\eqref{BFKL} as before.  We will later use the resummed
expressions shown in~\eqref{eq:Lnlo} to obtain explicit results.

In our calculation, we first perform the ${\bm x}_2$--integral to eliminate
the ${\bm q}'$ dependence and again use shifts and sign reflections in ${\bm
  q}$ to cast the result in the form
\begin{align}
  \label{eq:nonforward-abstract-target}
  \frac{\partial}{\partial Y} & \int\!\frac{d^2k}{(2\pi)^2}
  \frac{d^2l}{(2\pi)^2} 
  \left(e^{i {\bm l}\cdot{\bm x}_0}+e^{i {\bm l}{\bm x}_1}
  -e^{i{\bm k}\cdot{\bm x}_{01}+i{\bm l}\cdot \frac{{\bm x}_0+{\bm x}_1}2} 
  -e^{-i{\bm k}\cdot{\bm x}_{01}+i{\bm l}\cdot \frac{{\bm x}_0+{\bm x}_1}2} 
    \right) 
  G_{{\bm k}+\frac{{\bm l}}2, {\bm k}-\frac{{\bm l}}2 }(Y) 
\notag \\ = &
  \int\!\frac{d^2k}{(2\pi)^2} \frac{d^2l}{(2\pi)^2}
  \frac{N_c}{2\pi^2}\int d^2 q\ 
\ 
\left\{
 \left(e^{i {\bm l}\cdot{\bm x}_0}+e^{i {\bm l}{\bm x}_1}
  -e^{i{\bm q}\cdot{\bm x}_{01}+i{\bm l}\cdot \frac{{\bm x}_0+{\bm x}_1}2} 
  -e^{-i{\bm q}\cdot{\bm x}_{01}+i{\bm l}\cdot \frac{{\bm x}_0+{\bm x}_1}2} 
    \right)
  K_{\text{real}}({\bm q},{\bm k},{\bm l})
\right . \notag \\ & \left. \hspace{1cm}
- \left(e^{i {\bm l}\cdot{\bm x}_0}+e^{i {\bm l}{\bm x}_1}
  -e^{i{\bm k}\cdot{\bm x}_{01}+i{\bm l}\cdot \frac{{\bm x}_0+{\bm x}_1}2} 
  -e^{-i{\bm k}\cdot{\bm x}_{01}+i{\bm l}\cdot \frac{{\bm x}_0+{\bm x}_1}2} 
    \right)
K_{\text{virt}}({\bm k},{\bm q},{\bm l})
\right\}   G_{{\bm k}+\frac{{\bm l}}2, {\bm k}-\frac{{\bm l}}2 }(Y).
\end{align}
The kernels can again be expressed in terms of $L_{{\bm q},{\bm q}'}$ as
before in the forward case\footnote{Note that
  in~\eqref{eq:nonforward-abstract-target} we may reverse the sign of $\bm q$
  in $K_{\text{real}}({\bm q},{\bm k},{\bm l})$ on a term by term basis. The
  form chosen here reduces the number of steps needed to arrive at our final
  result~\eqref{eq:BK-nonforward-BFKL-Lotterlike}.}.  We find
\begin{subequations}
\label{eq:nonforward-kernels}  
\begin{align}
  K_{\text{real}}({\bm k},{\bm q},{\bm l}) &
  =
L_{{\bm k}-\frac{\bm l}2,-{\bm q}+{\bm k}}
-L_{{\bm k}-\frac{\bm l}2,{\bm k}+\frac{\bm l}2}
-L_{{\bm q}-{\bm k},-{\bm k}+\frac{\bm l}2}
+L_{-{\bm k}-\frac{\bm l}2,{\bm q}-{\bm k}}
-L_{-{\bm k}-\frac{\bm l}2,-{\bm k}+\frac{\bm l}2}
+L_{-{\bm q}+{\bm k},{\bm k}+\frac{\bm l}2},
\\ 
  K_{\text{virt}}({\bm k},{\bm q},{\bm l}) 
&
  =L_{{\bm q},-{\bm k}+\frac{\bm l}2+{\bm q}}
  +L_{{\bm q},{\bm k}+\frac{\bm l}2+{\bm q}}.
\end{align}
\end{subequations}
In these expressions we have already anticipated the swap of ${\bm k}$ and
${\bm q}$ in the real term to facilitate the last step, in which we read off
the equation for $G_{{\bm k}+\frac{{\bm l}}2, {\bm k}-\frac{{\bm l}}2 }(Y)$ by
equating the integrands:
\begin{align}
  \label{eq:nonforward-BFKL-abstract}
  \frac{\partial \, G_{{\bm k}+\frac{{\bm l}}2, {\bm k}-\frac{{\bm
          l}}2 }(Y)}{\partial Y} = & \frac{N_c}{2\pi^2}\int d^2 q\ 
  \Big\{ K_{\text{real}}({\bm k},{\bm q},{\bm l}) G_{{\bm
      q}+\frac{{\bm l}}2, {\bm q}-\frac{{\bm l}}2 }(Y)
- K_{\text{virt}}({\bm k},{\bm q},{\bm l})
  G_{{\bm k}+\frac{{\bm l}}2, {\bm k}-\frac{{\bm l}}2 }(Y)
\Big\}
\end{align}
Last, we insert the explicit expressions of Eq.~\eqref{eq:Lnlo}
into~\eqref{eq:nonforward-kernels} and repeat the arguments of the
forward case to again cancel the antisymmetric parts. Anticipating a
comparison with Levin and Braun~\cite{Levin:1994di, Braun:1994mw} we
introduce the notation
\begin{align}
  \label{eq:etadef}
  \eta\left({\bm k}^2\right) := 
  \frac{{\bm k}^2}{\alpha_s\left({\bm k}^2\,e^{-5/3}\right)}
\end{align}
and change notation by shifting both ${\bm k}$ and ${\bm q}$ by $-{\bm l}/2$.
Note that~\cite{Levin:1994di, Braun:1994mw} do not specify the renormalization
scheme and hence do not keep track of the factors $e^{-5/3}$ we take care to
include here.\footnote{The factor simply enters the definition of
  $\Lambda_{\text{QCD}}$ and emerges directly from our calculations in
  dimensional regularization. This is entirely independent from the issue of
  separation scheme dependence as discussed in Section~\ref{momNLO}.}

With these conventions, the final equation reads
\begin{align}
  \label{eq:BK-nonforward-BFKL-Lotterlike}
  \frac{\partial \, G_{{\bm k}, {\bm k}-{\bm l} }(Y)}{\partial Y} = &
  \frac{N_c}{2\pi^2} \int d^2q \ K_{\text{real}}({\bm k},{\bm q},{\bm
    l}) \ G_{{\bm q}, {\bm q}-{\bm l} }(Y)
-\left[\beta\left(\left({\bm k}-{\bm l}\right)^2\right)
+      \beta\left({\bm k}^2\right)
\right]G_{{\bm k}, {\bm k}-{\bm l} }(Y)
\end{align}
where the real kernel takes the explicit form
\begin{align}
  \label{eq:nonforward-real-final}
 K_{\text{real}}({\bm k},{\bm q},{\bm l}) =
\frac{\eta\left(\left({\bm q}-\bm l\right)^2\right)}{
  \eta\left(\left({\bm k}-\bm l\right)^2\right) 
  \eta\left((\left({\bm k}-\bm q\right)^2\right)}
+\frac{\eta\left({\bm q}^2\right)}{
  \eta\left({\bm k}^2\right) 
  \eta\left(\left({\bm k}-\bm q\right)^2\right)}
-\frac{\eta\left({\bm l}^2\right)}{
  \eta\left({\bm k}^2\right)
  \eta\left(\left({\bm k}-\bm l\right)^2\right)}
\end{align}
and $\beta\left({\bm k}^2\right)$ denotes the 
gluon trajectory defined as
\begin{align}
  \label{eq:gluon-traj-rep}
  \beta\left({\bm k}^2\right)
=
\frac{N_c}{2\pi^2}\int d^2q  \
\frac12\frac{\eta\left({\bm k}^2\right)}{
 \eta\left({\bm q}^2\right) 
  \eta\left(\left({\bm k}-\bm q\right)^2\right)}
\ .
\end{align}
We note that all individual contributions are composed of triumvirate
structures. In the forward limit, this reduces to what we found earlier
in~\eqref{eq:ian-G-eq} with $\Tilde N(k,Y) \approx G_{\bm k,\bm k}(Y)$.

We now wish to compare to~\cite{Levin:1994di, Braun:1994mw}, without loosing
the connection with the bootstrap condition used in their argument. This
prevents us from taking the forward limit (which we have already shown at the
end of Sec.~\ref{sec:resumm-bubble-diagr-forward}). Instead, we identify the
quantity used in~\cite{Levin:1994di, Braun:1994mw} by amputating the external,
coupling--resummed gluon legs from the dipole scattering amplitude $G_{{\bm
    k},{\bm k}'}(Y)$: we separate out a factor $\frac{\alpha_s\left({\bm
      k}^2\right)}{{\bm k}^2} \frac{\alpha_s\left(\left({\bm
        k}'\right)^2\right)}{{\bm k}'^2}$ according to
\begin{align}
  \label{eq:amputation}
  G_{{\bm k},{\bm k}'}(Y) =
  \parbox{3cm}{\includegraphics[width=3cm]{Gdef-mom-cut}} =
  \frac{\alpha_s\left({\bm k}^2\,e^{-5/3}\right)}{{\bm k}^2}
  \frac{\alpha_s\left({\bm k}'^2\,e^{-5/3}\right)}{{\bm k}'^2} \ 
  \varphi_{{\bm k},{\bm k}'}(Y) \ .
\end{align}
Generically, such modifications leave the gluon Regge trajectories
unaffected and rescale only the real kernel. Here, the real kernel is
modified to
\begin{align}
  \label{eq:explcit-Levin-nonforward-real}
  \Tilde K_{\text{real}}({\bm k},{\bm q},{\bm l}) = &
  K_{\text{real}}({\bm q},{\bm k},{\bm l}) \ \frac{ \eta\left({\bm
        k}^2\right)\eta\left(\left({\bm k}-{\bm l}\right)^2\right) }{
    \eta\left({\bm q}^2\right)\eta\left(\left({\bm q}-{\bm
          l}\right)^2\right) }
  \notag \\
  = & \frac{ \eta\left(\left({\bm k}-\bm l\right)^2\right) }{
    \eta\left(\left({\bm q}-\bm l\right)^2\right)
    \eta\left((\left({\bm k}-\bm q\right)^2\right) } +\frac{
    \eta\left({\bm k}^2\right) }{ \eta\left({\bm q}^2\right)
    \eta\left(\left({\bm k}-\bm q\right)^2\right) } -\frac{
    \eta\left({\bm l}^2\right) }{ \eta\left({\bm q}^2\right)
    \eta\left(\left({\bm k}-\bm l\right)^2\right)} \ .
\end{align}
This turns Eq.~\eqref{eq:BK-nonforward-BFKL-Lotterlike} into
\begin{align}
  \label{eq:BK-nonforward-BFKL-Lotterlike2}
  \frac{\partial \varphi_{{\bm k}, {\bm k}-{\bm l} }(Y)}{\partial Y} =
  & \frac{N_c}{2\pi^2} \int d^2q\ 
  \Tilde K_{\text{real}}({\bm k},{\bm q},{\bm l}) 
  \ \varphi_{{\bm q}, {\bm q}-{\bm l} }(Y)
-\left[\beta\left(\left({\bm k}-{\bm l}\right)^2\right)
+      \beta\left({\bm k}^2\right)
\right]\varphi_{{\bm k}, {\bm k}-{\bm l} }(Y)
\ .
\end{align}
The structures in this equation correspond directly to (16)
of~\cite{Levin:1994di} or (4) of~\cite{Braun:1994mw}, which, as already
mentioned, were based on calculating the running coupling corrections to the
virtual contributions and \emph{postulating} a bootstrap condition to extend
the result to include corrections for the real contributions.

Our calculation now provides a derivation of this statement in the $s$-channel light cone perturbation theory formalism, previously used to derive the JIMWLK and BK equations, and gives a firm interpretation of the objects involved.


\section{Leading-$N_f$ contribution to the NLO BFKL pomeron intercept}
\label{NLOint}

To find the leading-$N_f$ contribution to the NLO BFKL intercept (in
the large-$N_c$ limit) we have to act on $\phi (k, Y) \sim k^{2 \,
  \lambda}$ with the NLO part of the kernel in \eq{NLO_BFKL}.
Employing \eq{measure} again and using the same notation as in Sect.
\ref{LOint} we write
\begin{align}\label{nloeig1}
  & \int d^2 q \, K^{\text{NLO}}_f ({\bm k}, {\bm q}) \, q^{2 \,
    \lambda} \, \equiv \notag \\ & \equiv \, \int d^2 q \, \left[
    \frac{2}{({\bm k} - {\bm q})^2} \, \ln \left( \frac{ ({\bm k} -
        {\bm q})^2}{\mu^2 } \right) \, q^{2 \, \lambda} - \frac{{\bm
        k}^2}{{\bm q}^2 \, ({\bm k} - {\bm q})^2} \, \ln \left(
      \frac{{\bm q}^2 \, ({\bm k} - {\bm q})^2}{{\bm k}^2 \, \mu^2 }
    \right) \, k^{2 \, \lambda} \right] \notag \\ & = \, 2 \pi \,
  \int\limits_0^\infty db \, b \, J_0 (b \, k) \,
  \int\limits_\Lambda^\infty dq \, q \, J_0 (b \, q) \,
  \int\limits_\Lambda^\infty dl \, l^{-1} \, J_0 (b \, l) \times \,
  \left[ 2 \, \ln \left( \frac{l^2}{\mu^2 } \right) \, q^{2 \,
      \lambda} - \frac{k^2}{q^2} \, \ln \left( \frac{q^2 \, l^2}{k^2
        \, \mu^2 } \right) \, k^{2 \, \lambda} \right]
\end{align}
Integrating over $q$ and $l$ with the help of formulae in Appendix
\ref{App2} yields
\begin{align}\label{nloeig2}
  \int d^2 q \, K^{\text{NLO}}_f ({\bm k}, {\bm q}) \, q^{2 \,
    \lambda} \, = \notag \\ = 2 \pi \, \int\limits_0^\infty db \, b \,
  J_0 (b \, k) \, \left( \psi (1) - \ln \frac{b \, \Lambda}{2} \right)
  \, \Bigg\{ 2^{2 \lambda + 2} \, b^{-2 \lambda -2} \, \frac{\Gamma
    \left( 1 + \lambda \right)}{\Gamma \left( - \lambda \right)} \,
  \left[\psi (1) - \ln \frac{b \, \mu^2}{2 \, \Lambda} \right] \notag
  \\ - 2 \, k^{2 \, \lambda + 2} \, \left( \psi (1) - \ln \frac{b \,
      \Lambda}{2} \right) \, \left( \psi (1) - \ln \frac{b \, k \, \mu
    }{2 \, \Lambda} \right) \Bigg\}.
\end{align}
Finally, using formulae in Appendix \ref{App2} to integrate
\eq{nloeig2} over $b$ we obtain
\begin{align}\label{nloeig3}
  \int d^2 q \, K^{\text{NLO}}_f ({\bm k}, {\bm q}) \, q^{2 \,
    \lambda} \, = \, 2 \pi \, k^{2 \, \lambda} \, \left\{ \frac{1}{2}
    \, \chi^2 (-\lambda) + \chi (-\lambda) \, \ln \frac{k^2}{\mu^2} +
    \frac{1}{2} \, \psi' (-\lambda) - \frac{1}{2} \, \psi'
    (1+\lambda)\right\}.
\end{align}
Including the prefactor of $\frac{\am^2 \, N_c \, N_f}{12 \, \pi^3}$
which we have been omitting above, combining (\ref{nloeig3}) with the
leading order contribution to the intercept (\ref{loeig3}) and
remembering that $\mu^2 = \mu_{\overline MS}^2 \, e^{5/3}$ yields
\begin{align}\label{int}
  & \frac{\am \, N_c}{\pi} \, \int d^2 q \, K^{\text{LO}} ({\bm k},
  {\bm q}) \, \left(\frac{q}{k}\right)^{2 \, \lambda} + \frac{\am^2 \,
    N_c \, N_f}{12 \, \pi^3} \, \int d^2 q \, K^{\text{NLO}}_f ({\bm
    k}, {\bm q}) \, \left(\frac{q}{k}\right)^{2 \, \lambda} \, =
  \notag \\ & = \frac{\am N_c}{\pi} \left\{ \chi (-\lambda) \left[ 1 -
      \am \, \beta_2 \, \ln \frac{k^2}{\mu_{\overline{\text{MS}}}^2}
    \right] - \frac{\am N_f}{12 \, \pi} \left[-\chi^2 (-\lambda) -
      \psi' (-\lambda) + \psi' (1+\lambda) + \frac{10}{3} \, \chi
      (-\lambda) \right] \right\}
\end{align}
where we have replaced $N_f \rightarrow - 6 \pi \beta_2$ in front of the
logarithm to underline the fact that this separation of terms has often been
interpreted a separation of running coupling and conformal contributions.

From our earlier considerations in Sec.~\ref{momNLO}, we now know that such a
separation is artificial. In fact the discussion there and in
Sec.~\ref{sec:resumm-bubble-diagr-forward} as well as
Sec.~\ref{sec:gener-nonf-bfkl} allow us to take the extreme position and
assign all contributions listed here to running coupling effects. To
facilitate comparison with~\cite{Camici:1997ta,Fadin:1998py}, we still
introduce the notation $\Delta (\lambda)$ for the last term in Eq.~\eqref{int}
by
\begin{align}
  \frac{\am \, N_c}{\pi} \, \int d^2 q \, K^{\text{LO}} ({\bm k}, {\bm
    q}) \, \left(\frac{q}{k}\right)^{2 \, \lambda} + & \frac{\am^2 \,
    N_c \, N_f}{12 \, \pi^3} \, \int d^2 q \, K^{\text{NLO}}_f ({\bm
    k}, {\bm q}) \, \left(\frac{q}{k}\right)^{2 \, \lambda} \, =
  \notag \\ = \, \frac{\am \, N_c}{\pi} \, \Bigg\{ \chi (-\lambda) \,
  \bigg[ 1 - \am \, \beta_2 \, & \ln
  \frac{k^2}{\mu_{\overline{\text{MS}}}^2} \bigg] - \frac{\am \,
    N_f}{12 \, \pi} \, \Delta (\lambda) \Bigg\}
\ .
\end{align}
From~\eq{int} we have
\begin{align}\label{intd}
  \Delta (\lambda) \, = \, - \chi^2 (-\lambda) - \psi' (-\lambda) +
  \psi' (1+\lambda) + \frac{10}{3} \, \chi (-\lambda).
\end{align}

To compare our result with the result of the NLO calculation of Fadin
and Lipatov \cite{Fadin:1998py} and of Camici and Ciafaloni
\cite{Camici:1997ta} we note that in \cite{Fadin:1998py,Camici:1997ta}
instead of the power $\lambda$ the authors used the power $\gamma -1$,
such that $\lambda = \gamma -1$. We thus first rewrite \eq{intd} in
terms of $\gamma$
\begin{align}\label{int_gamma}
  \Delta (\gamma) \, = \, - \chi^2 (\gamma) - \psi' (1-\gamma) + \psi'
  (\gamma) + \frac{10}{3} \, \chi (\gamma).
\end{align}

The intercept of \eq{int_gamma} exactly agrees with the result of
Fadin and Lipatov (see Eqs.  (14) and (12) in \cite{Fadin:1998py}) and
with the result of Camici and Ciafaloni (see Eq. (4.6) in
\cite{Camici:1997ta})! We have thus provided an independent
cross-check of those earlier results~\cite{Camici:1997ta,Fadin:1998py}.

It is important to note that the agreement between the intercepts
of~\cite{Camici:1997ta,Fadin:1998py} and the one in~\eq{int_gamma} depends
crucially on the fact that both intercepts were calculated for the same
observable --- the unintegrated gluon distribution. We also stress that the
factor of $\as (k^2)$ in the definition of the unintegrated gluon distribution
in~\eq{phi} plays an important role in obtaining the correct intercept.


\section{Comparison with the NLO intercept for the evolution of dipole 
  amplitude calculated in~\cite{Bal:2006}}
\label{Ian}

Another leading-$N_f$ NLO BFKL intercept we should compare with is due to a
transverse coordinate space calculation performed by Balitsky~\cite{Bal:2006}.
The result of~\cite{Bal:2006}, expressed in terms of the same power $\gamma$
as defined in~\cite{Camici:1997ta,Fadin:1998py} reads (see Eq. (44) in
\cite{Bal:2006})
\begin{align}\label{Bal}
  \Delta_{\text{B}} (\gamma) \, = \, \chi^2 (\gamma) + \psi'
  (1-\gamma) - \psi' (\gamma) - \frac{4}{1-\gamma} \, \chi (\gamma) +
  \frac{10}{3} \, \chi (\gamma).
\end{align}
We will demonstrate below that this expression can be translated into the
corresponding (uniquely defined) contribution in the results
of~\cite{Camici:1997ta,Fadin:1998py} and our earlier result~\eq{int_gamma}, by
taking into account that (i) $\Delta_{\text{B}} (\gamma)$ was obtained using
both kernels and eigenfunctions in transverse coordinate space and that (ii)
$\Delta_{\text{B}} (\gamma)$ is the intercept for a different observable ---
the forward scattering amplitude for a dipole on the target, while $\Delta
(\gamma)$ was calculated for the unintegrated gluon distribution in momentum
space.

We will, therefore, show that after a Fourier transform relating $\phi (k,Y)$
to $N (x_{01}, Y)$, our intercept from \eq{int_gamma} is translates into the
intercept $\Delta_{\text{B}} (\gamma)$ from \eq{Bal} obtained in
\cite{Bal:2006}. Combining \eq{bindep} with \eq{FT} we write
\begin{align}
  N (x_{01}, Y) \, = \, \frac{2 \, \pi}{N_c} \, \int d^2 k \, (1 -
  e^{i {\bm k} \cdot {\bm x}_{01}}) \, \frac{\as (k^2) \, \phi (k,
    Y)}{k^2}.
\end{align}
Therefore, if we put $\as (k^2) \, \phi (k, Y) \, = \, k^{2 \,
  \lambda}$ then we would obtain
\begin{align}
  N (x_{01}, Y) \, = \, - \frac{(2 \, \pi)^2}{N_c} \, 2^{2 \, \lambda
    -1} \, x_{01}^{- 2 \, \lambda} \, \frac{\Gamma (\lambda)}{\Gamma
    (1-\lambda)}.
\end{align}
Using \eq{NLO_BFKL_1} and the steps which led to it we write
\begin{align}\label{bal1}
  \am^2 \, \int d^2 x_2 \, K^{\text{NLO}}_f ({\bm x}_0, {\bm x}_1 ;
  {\bm x}_2) \, \left[ N ({\bm x}_{0}, {\bm x}_2, Y) + N ({\bm x}_{2},
    {\bm x}_1, Y) - N ({\bm x}_{0}, {\bm x}_1, Y) \right] \nonumber \\
  \, = \, \frac{2 \, \pi}{N_c} \, \int \frac{d^2 k}{{\bm k}^2} \, (1 -
  e^{i {\bm k} \cdot {\bm x}_{01}}) \, \frac{\am^2 \, N_c \, N_f}{12
    \, \pi^3} \, \int d^2 q \, \left[ K^{\text{NLO}}_f ({\bm k}, {\bm
      q}) + \frac{2}{({\bm k} - {\bm q})^2} \, \ln \frac{{\bm
        k}^2}{{\bm q}^2}\right] \, \as (q^2) \, \phi (q, Y).
\end{align}
Therefore, for $N (x_{01}, Y) = x_{01}^{- 2 \, \lambda}$,
\begin{align}\label{bal2}
  \am^2 \, K_f^{\text{NLO}} \, \otimes \, x_{01}^{- 2 \, \lambda} \, =
  \, - \frac{2^{-2 \, \lambda +1}}{2 \, \pi} \, \frac{\Gamma
    (1-\lambda)}{\Gamma (\lambda)}\int \frac{d^2 k}{{\bm k}^2} \, (1 -
  e^{i {\bm k} \cdot {\bm x}_{01}}) \nonumber \\ \times \, \frac{\am^2
    \, N_c \, N_f}{12 \, \pi^3} \, \int d^2 q \, \left[
    K^{\text{NLO}}_f ({\bm k}, {\bm q}) + \frac{2}{({\bm k} - {\bm
        q})^2} \, \ln \frac{{\bm k}^2}{{\bm q}^2}\right] \, q^{2 \,
    \lambda},
\end{align}
where we have abbreviated the action of the NLO kernel on the left
hand side of \eq{bal1}. To perform the part of $q$-integral involving
$K^{\text{NLO}}_f ({\bm k}, {\bm q})$ we will employ \eq{nloeig3}. The
rest of the $q$-integral is
\begin{align}
  \int d^2 q \, \frac{2}{({\bm k} - {\bm q})^2} \, \ln \frac{{\bm
      k}^2}{{\bm q}^2} \, q^{2 \, \lambda} \, = \, 2 \, \pi \,
  \left[\psi' (1+ \lambda) - \psi' (- \lambda) \right] \, k^{2 \,
    \lambda} ,
\end{align}
which can be derived using \eq{measure} and the formulae in Appendix
\ref{App2}. Combining this with \eq{nloeig3} yields
\begin{align}\label{bal3}
  \am^2 \, K_f^{\text{NLO}} \, \otimes \, x_{01}^{- 2 \, \lambda} \, =
  \, - 2^{-2 \, \lambda +1} \, \frac{\Gamma (1-\lambda)}{\Gamma
    (\lambda)}\int \frac{d^2 k}{{\bm k}^2} \, (1 - e^{i {\bm k} \cdot
    {\bm x}_{01}}) \nonumber \\ \times \, \frac{\am^2 \, N_c \,
    N_f}{12 \, \pi^3} \, k^{2 \, \lambda} \, \left\{ \frac{1}{2} \,
    \chi^2 (-\lambda) + \chi (-\lambda) \, \ln \frac{k^2}{\mu^2} -
    \frac{1}{2} \, \psi' (-\lambda) + \frac{1}{2} \, \psi'
    (1+\lambda)\right\}.
\end{align}
Performing the integration over ${\bm k}$ we obtain
\begin{align}\label{bal4}
  \am^2 \, K_f^{\text{NLO}} \, \otimes \, x_{01}^{- 2 \, \lambda} \, =
  \, \frac{\am^2 \, N_c \, N_f}{6 \, \pi^2} \, \Bigg\{ \frac{1}{2} \,
  \chi^2 (-\lambda) + \chi (-\lambda) \, \left[ \ln \frac{4 \, e^{- 2
        \, \gamma}}{x_{01}^2 \, \mu^2} - \chi (-\lambda) -
    \frac{2}{\lambda} \right] \notag \\ - \, \frac{1}{2} \, \psi'
  (-\lambda) + \frac{1}{2} \, \psi' (1+\lambda) \Bigg\} \, x_{01}^{- 2
    \, \lambda},
\end{align}
where $\gamma = - \psi (1)$ is the Euler's constant. Since the LO BFKL
kernel acting on a power in transverse coordinate space gives the
usual BFKL eigenvalue
\begin{align}
  \am \, K^{\text{LO}} \, \otimes \, x_{01}^{- 2 \, \lambda} \, = \,
  \frac{\am \, N_c}{\pi} \, \chi (-\lambda) \, x_{01}^{- 2 \, \lambda}
\end{align}
we write
\begin{align}\label{bal5}
  \left[ \am \, K^{\text{LO}} + \am^2 \, K_f^{\text{NLO}} \right]
  \otimes \, x_{01}^{- 2 \, \lambda} \, = \, \frac{\am \, N_c}{\pi} \,
  \Bigg\{ \chi (-\lambda) \, \left[ 1 - \am \, \beta_2 \, \ln \frac{4
      \, e^{- 2 \, \gamma}}{x_{01}^2 \, \mu^2_{\overline{\text{MS}}}}
  \right] - \frac{\am \, N_f}{12 \, \pi} \, \Delta_B (\lambda) \Bigg\}
  \, x_{01}^{- 2 \, \lambda}
\end{align}
with
\begin{align}\label{bal6}
  \Delta_B (\lambda) \, = \, \chi^2 (-\lambda) + \frac{4}{\lambda} \,
  \chi (-\lambda) + \psi' (-\lambda) - \psi' (1+\lambda) +
  \frac{10}{3} \, \chi (-\lambda).
\end{align}
Here one might worry that this value of $\Delta_B (\lambda)$ depends
on our choice of the constant under the logarithm in \eq{bal5}.
However this choice is not arbitrary and is consistent with the
constant obtained under the logarithm of the coordinate space running
coupling corrections resummed to all orders in \cite{HW-YK:2006}. (In
\cite{HW-YK:2006} we would also have a factor of $e^{-5/3}$ under the
running coupling logarithm. Here, following the convention of
\cite{Camici:1997ta,Fadin:1998py,Bal:2006} we have chosen to place this
contribution separately: it leads to the $\frac{10}{3} \, \chi
(-\lambda)$ term in \eq{bal6}.)

Replacing $\lambda = \gamma -1$ reduces \eq{bal6} to \eq{Bal}! We have
thus shown that our intercept (\ref{int_gamma}) and the intercept in
\eq{Bal} found in \cite{Bal:2006} are consistent with each other. The
difference between them is due to the fact that they are calculated
for two different observables.


\section{Conclusions}

In this paper we have studied the implications for the linear BFKL
evolution equation of including running coupling corrections into the
non-linear JIMWLK and BK evolution equations performed in
\cite{HW-YK:2006}. In particular we have derived the BFKL equation
including running coupling corrections to all orders and calculated
the leading-$N_f$ NLO BFKL pomeron intercept. Until now the form of
the BFKL equation with {\em all orders resummed} running coupling
corrections was not known. What existed was a conjecture by Braun and
by Levin~\cite{Braun:1994mw,Levin:1994di} based on an interesting
assumption that bootstrap equations hold even with running coupling
corrections included.  We have shown that this conjecture is in fact
accurate, though for a slightly different observable than suggested
originally. Our results were derived using the $s$-channel language of
light cone perturbation theory \cite{Lepage:1980fj,Brodsky:1997de}. Up
to now the NLO BFKL intercept had only been calculated either by using
the standard Feynman perturbation theory
in~\cite{Camici:1997ta,Fadin:1998py} and or by employing the
background field method~\cite{Bal:2006}. Our calculation provides an
independent check of the intercept found
in~\cite{Camici:1997ta,Fadin:1998py} and connects it to the one
obtained in~\cite{Bal:2006}.


Our paper has two main results. First of all, we have obtained the
BFKL equation for the unintegrated gluon distribution including
running coupling corrections resummed to all orders.  The equation is
given by (\ref{rc_BFKL}). After a redefinition of the unintegrated
gluon distribution shown in \eq{tphi},~\eq{rc_BFKL} leads
to~\eq{rc_BFKL_Genya}, which was conjectured by Braun and by Levin
in~\cite{Braun:1994mw,Levin:1994di} by postulating the bootstrap
condition in the running coupling case. We have thus shown that the
conjecture of~\cite{Braun:1994mw,Levin:1994di} for the BFKL equation
with the running coupling corrections is correct, though for a
slightly non-traditional definition of the unintegrated gluon
distribution~(\ref{tphi}). We have clarified that this result is based
on a complete absorption of all $\as \, N_f$ corrections into the
running of the coupling in the BFKL equation (as in~\eq{rc_BFKL}).
This is one of many possible constructions as explained in
Sec.~\ref{momNLO}, all others would separate off explicit
$N_f$--contributions that are not interpreted as running coupling
contributions. The result~\eq{rc_BFKL} is formulated in the form of a
``triumvirate'' of the couplings~\cite{HW-YK:2006}.

Our second result is an independent check of the leading-$N_f$
contribution to the NLO BFKL intercept (expanded to strictly NLO,
without any resummations), which we performed in Sect.~\ref{NLOint}.
Our intercept, obtained for the evolution of the unintegrated gluon
distribution function, is given by~\eq{int_gamma} and completely
agrees with the results of Camici and Ciafaloni~\cite{Camici:1997ta}
and of Fadin and Lipatov~\cite{Fadin:1998py}.  The NLO intercept
appears to strongly depend on the physical observable: we demonstrate
that by calculating the leading-$N_f$ NLO BFKL intercept for the
dipole scattering amplitude, starting from that of the unintegrated
gluon distribution. Our result is given in~\eq{bal6} and agrees with
the result of Balitsky~\cite{Bal:2006}. The difference in the
expressions~\eqref{intd} and~\eqref{bal6} is fully explained by the
fact that the two intercepts refer to different observables.


\section*{Acknowledgments} 

We would like to thank Ian Balitsky for many informative discussions
on the subject.

Yu.K. would like to thank the Department of Energy's Institute for
Nuclear Theory at the University of Washington for its hospitality.
The work of Yu.K. is supported in part by the U.S.  Department of
Energy under Grant No.  DE-FG02-05ER41377.


\appendix

\renewcommand{\theequation}{A\arabic{equation}}
  \setcounter{equation}{0}
\section{Useful formulae}
\label{App2}

Here we list some useful mathematical formulae. The main formula we
need above is
\begin{align}\label{master}
  \int\limits_0^\infty dl \, l^{2 \lambda + 1} \, J_0 (b \, l) \, = \,
  2^{2 \lambda + 1} \, b^{-2 \lambda -2} \, \frac{\Gamma \left( 1 +
      \lambda \right)}{\Gamma \left( - \lambda \right)}.
\end{align}
Using \eq{master} one can derive the following useful results (here
$b>0$):
\begin{align}\label{B2}
  \int\limits_\Lambda^\infty dl \, l^{-1} \, J_0 (b \, l) \, = \, \psi
  (1) - \ln \frac{b \, \Lambda}{2},
\end{align}
\begin{align}\label{B3}
  \int\limits_\Lambda^\infty dl \, l^{-1} \, J_0 (b \, l) \, \ln l^2 =
  \, \left( \psi (1) - \ln \frac{b \, \Lambda}{2} \right) \, \left(
    \psi (1) - \ln \frac{b}{2 \, \Lambda} \right),
\end{align}
\begin{align}\label{B4}
  \int\limits_0^\infty dl \, l \, J_0 (b \, l) \, = \, 0,
\end{align}
\begin{align}\label{B5}
  \int\limits_0^\infty dl \, l \, J_0 (b \, l) \, \ln l \, = \, - \frac{1}{b^2},
\end{align}
\begin{align}\label{B6}
  \int\limits_0^\infty dl \, l \, J_0 (b \, l) \, \ln^2 l \, = \,
  \frac{2}{b^2} \, \left[ \ln \left(\frac{b}{2}\right) - \psi (1) \right],
\end{align}
\begin{align}\label{B7}
  \int\limits_0^\infty dl \, l \, J_0 (b \, l) \, \ln^3 l \, = \, -
  \frac{3}{b^2} \, \left[ \ln \left(\frac{b}{2}\right) - \psi (1)
  \right]^2.
\end{align}


\renewcommand{\theequation}{B\arabic{equation}}
  \setcounter{equation}{0}
\section{Evaluating \eq{K1Bal2}}
\label{App1}

Our goal here is to perform the $\bm k$-integration in \eq{K1Bal2}. We
begin with the first term in the curly brackets in \eq{K1Bal2}.
Performing the $\alpha$-integral first we obtain
\begin{align}\label{A1}
  4 \, N_f \, \int \frac{d^2 q}{(2\pi)^2}\frac{d^2
    q'}{(2\pi)^2}\frac{d^d k}{(2\pi)^d} \ e^{ -i {\bm q} \cdot ({\bm
      z}_1 - {\bm x}_0) + i {\bm q} \cdot ({\bm z}_1 - {\bm x}_1)} \,
  \frac{1}{{\bm q}^2 \, {\bm q}^{\prime 2}} \, \frac{({\bm q}- {\bm k})
    \cdot ({\bm q}' - {\bm k})}{({\bm q}- {\bm k})^2 - ({\bm q}' -
    {\bm k})^2} \, \ln \frac{({\bm q}- {\bm k})^2}{({\bm q}' - {\bm
      k})^2}.
\end{align}
\eq{A1} can be rewritten as
\begin{align}\label{A2}
  4 \, N_f \, \int \frac{d^2 q}{(2\pi)^2}\frac{d^2
    q'}{(2\pi)^2}\frac{d^d k}{(2\pi)^d} \ e^{ -i {\bm q} \cdot ({\bm
      z}_1 - {\bm x}_0) + i {\bm q} \cdot ({\bm z}_1 - {\bm x}_1)} \,
  \frac{1}{{\bm q}^2 \, {\bm q}^{\prime 2}} \, ({\bm q}- {\bm k}) \cdot
  ({\bm q}' - {\bm k}) \notag \\ \times \, \int\limits_0^1 \, d\beta \, \frac{1}{({\bm
      q}- {\bm k})^2 \, (1-\beta)+ ({\bm q}' - {\bm k})^2 \, \beta}.
\end{align}
Defining a new integration variable ${\tilde{\bm k}} = {\bm k} -
(1-\beta) \, {\bm q} - \beta \, {\bm q}'$ we get
\begin{align}\label{A3}
  4 \, N_f \, \int \frac{d^2 q}{(2\pi)^2}\frac{d^2 q'}{(2\pi)^2} \ e^{
    -i {\bm q} \cdot ({\bm z}_1 - {\bm x}_0) + i {\bm q} \cdot ({\bm
      z}_1 - {\bm x}_1)} \, \frac{1}{{\bm q}^2 \, {\bm q}^{\prime 2}} \,
  \int\limits_0^1 \, d\beta \, \int \frac{d^d {\tilde k}}{(2\pi)^d} \,
  \frac{{\tilde{\bm k}}^2 - \beta \, (1-\beta) \, ({\bm q} - {\bm
      q}')^2}{{\tilde{\bm k}}^2 + \beta \, (1-\beta) \, ({\bm q} -
    {\bm q}')^2},
\end{align}
where we dropped the terms linear in ${\tilde{\bm k}}$ in the
numerator as they vanish after angular integration. Performing the
${\tilde{\bm k}}$-integral yields
\begin{align}\label{A4}
  -8 \, N_f \, \int \frac{d^2 q}{(2\pi)^2}\frac{d^2 q'}{(2\pi)^2} \ 
  e^{ -i {\bm q} \cdot ({\bm z}_1 - {\bm x}_0) + i {\bm q} \cdot ({\bm
      z}_1 - {\bm x}_1)} \, \frac{1}{{\bm q}^2 \, {\bm q}^{\prime 2}} \,
  \int\limits_0^1 \, d\beta \ \frac{\Gamma \left( 1 - \frac{d}{2}
    \right)}{(4 \, \pi)^{d/2}} \, \left[ \beta \, (1-\beta) \, ({\bm q} -
    {\bm q}')^2 \right]^{d/2}.
\end{align}
Inserting $d = 2 - \epsilon$, expanding the expression in powers of
$\epsilon$, replacing $1/\epsilon$ with $\ln \mu_{MS}$ and integrating
over $\beta$ yields
\begin{align}\label{A5}
  \frac{N_f}{3 \, \pi} \, \int \frac{d^2 q}{(2\pi)^2}\frac{d^2
    q'}{(2\pi)^2} \ e^{ -i {\bm q} \cdot ({\bm z}_1 - {\bm x}_0) + i
    {\bm q} \cdot ({\bm z}_1 - {\bm x}_1)} \, \frac{1}{{\bm q}^2 \,
    {\bm q}^{\prime 2}} \, ({\bm q} - {\bm q}')^2 \, \ln \frac{({\bm q} -
    {\bm q}')^2 \, e^{-5/3}}{\mu_{\overline{\text{MS}}}^2}
\end{align}
where $\mu_{\overline{\text{MS}}}^2 = \mu_{MS}^2 \, 4 \pi \,
e^{-\gamma}$. \eq{A5} gives us the first term in the curly brackets of
\eq{K1Bal3}.

The last term in the curly brackets of \eq{K1Bal2} gives us zero after
performing a dimensionally regularized $\bm k$-integral. We are left
only with the second and the third terms in the curly brackets of
\eq{K1Bal2}, which are analogous to each other. Here we will show how
to do the $\bm k$-integration in the second term only: the integral in
the third term can be easily done in the the same way.

Defining ${\tilde{\bm k}} = {\bm k} - (1-\alpha) \, {\bm q}$ the
second term in the curly brackets of \eq{K1Bal2} can be written as
\begin{align}\label{A6}
  - 4 \, N_f \, \int \frac{d^2 q}{(2\pi)^2}\frac{d^2 q'}{(2\pi)^2} \ 
  e^{ -i {\bm q} \cdot ({\bm z}_1 - {\bm x}_0) + i {\bm q} \cdot ({\bm
      z}_1 - {\bm x}_1)} \, \frac{1}{{\bm q}^2 \, {\bm q}^{\prime 2}} \,
  \int\limits_0^1 \, d\alpha \, \int \frac{d^d {\tilde k}}{(2\pi)^d}
  \, \frac{{\tilde{\bm k}}^2 - \alpha \, (1-\alpha) \, {\bm
      q}^2}{{\tilde{\bm k}}^2 + \alpha \, (1-\alpha) \, {\bm q}^2},
\end{align}
which, repeating the above steps which led from \eq{A3} to \eq{A5},
can be recast into
\begin{align}\label{A7}
  - \frac{N_f}{3 \, \pi} \, \int \frac{d^2 q}{(2\pi)^2}\frac{d^2
    q'}{(2\pi)^2} \ e^{ -i {\bm q} \cdot ({\bm z}_1 - {\bm x}_0) + i
    {\bm q} \cdot ({\bm z}_1 - {\bm x}_1)} \, \frac{1}{{\bm q}^{\prime 2}}
  \, \ln \frac{{\bm q}^2 \, e^{-5/3}}{\mu_{\overline{\text{MS}}}^2}.
\end{align}
This is exactly the second term in the curly brackets of \eq{K1Bal3}.
The third term is done by analogy with the second one.


\providecommand{\href}[2]{#2}\begingroup\raggedright\endgroup


\end{document}